\documentclass[12pt]{article}
\pdfoutput=1
\usepackage[DIV13]{typearea}
\usepackage{indentfirst}
\usepackage{amsmath}
\usepackage{mathrsfs}
\usepackage{amssymb}
\usepackage{bbm}
\usepackage{epsfig}
\usepackage{graphicx}
\usepackage{subfigure}
\usepackage{slashed}
\usepackage{multicol}
\usepackage[usenames,dvipsnames]{color}
\usepackage{cite}
\RequirePackage[colorlinks=true,urlcolor=blue,anchorcolor=blue,citecolor=blue,filecolor=blue,
               linkcolor=blue,menucolor=blue,linktocpage=true,pdfproducer=medialab]{hyperref}

\usepackage{cancel}
\newcommand{\email}[1]{\href{mailto:#1}{\tt #1}}

\numberwithin{equation}{section}
\newcommand{\LL}{\mathscr{L}}

\def\cD{{\cal D}}
\def\cF{{\cal F}}
\def\cO{{\cal O}}

\def\cX{{\cal X}}

\def\cy{{\bf y}}
\def\cY{{\bf Y}}
\def\Tr{{\rm Tr}}
\def\vep{\varepsilon}
\def\be{\begin{equation}}
\def\ee{\end{equation}}
\def\beq{\begin{equation}}
\def\eeq{\end{equation}}
\def\bc{\begin{center}}
\def\ec{\end{center}}
\def\bea{\begin{eqnarray}}
\def\eea{\end{eqnarray}}

\def\nn{\nonumber}

\newcommand{\mean}[1]{\langle#1\rangle}
\newcommand{\derp}{\partial}

\newcommand{\hc}{\text{h.c.}}
\newcommand{\unity}{\mathbbm{1}}
\newcommand{\diag}{{\rm{\bf diag}}}
\newcommand{\UH}{\mathbf{U}}

\newcommand{\TL}{\mathbf{T}}
\newcommand{\VL}{\mathbf{V}}
\newcommand{\DL}{\mathbf{D}}
\newcommand{\WL}{\mathbf{W}}

\newcommand{\tmn}{\sigma_{\mu\nu}}
\newcommand{\tmnp}{\sigma^{\mu\nu}}

\newcommand{\blue}[1]{\color{blue} #1 \color{black}}

\begin{document}
\begin{titlepage}
\vspace*{-1cm}
\phantom{hep-ph/***} 
{\flushleft
{\blue{FTUAM-12-116}}
\hfill{\blue{CERN-PH-TH/2012-337}}\\
{\blue{IFT-UAM/CSIC-12-115}}
\hfill{\blue{FLAVOUR(267104)-ERC-31}}\\
{\blue{DFPD-2012/TH/24}}}
\vskip 1.5cm
\begin{center}
\mathversion{bold}
{\LARGE\bf Flavour with a }\\[3mm]
{\LARGE\bf Light Dynamical ``Higgs Particle''}\\[3mm]
\mathversion{normal}
\vskip .3cm
\end{center}
\vskip 0.5  cm
\begin{center}
{\large R.~Alonso}~$^{a)}$,
{\large M.B.~Gavela}~$^{a,b)}$,\\[2mm]
{\large L.~Merlo}~$^{a,b,c)}$,
{\large S.~Rigolin}~$^{d)}$,
{\large and J.~Yepes}~$^{a)}$
\\
\vskip .7cm
{\footnotesize
$^{a)}$~
Departamento de F\'isica Te\'orica and Instituto de F\'{\i}sica Te\'orica, IFT-UAM/CSIC,\\
Universidad Aut\'onoma de Madrid, Cantoblanco, 28049, Madrid, Spain\\
\vskip .1cm
$^{b)}$~
CERN, Department of Physics, Theory Division CH-1211 Geneva 23, Switzerland\\
\vskip .1cm
$^{c)}$~
TUM Institute for Advanced Study, Technische Universit\"at M\"unchen, \\
Lichtenbergstrasse 2a, D-85748 Garching, Germany
\vskip .1cm
$^{d)}$~
Dipartimento di Fisica ``G.~Galilei'', Universit\`a di Padova and \\
INFN, Sezione di Padova, Via Marzolo~8, I-35131 Padua, Italy
\vskip .5cm
\begin{minipage}[l]{.9\textwidth}
\begin{center} 
\textit{E-mail:} 
\email{rodrigo.alonso@uam.es},
\email{belen.gavela@uam.es},
\email{luca.merlo@uam.es},
\email{stefano.rigolin@pd.infn.it},
\email{ju.yepes@uam.es}
\end{center}
\end{minipage}
}
\end{center}
\vskip 1cm
\begin{abstract}
The Higgs-fermion couplings are sensitive probes of possible new physics  behind  a stable light Higgs  particle. It is then essential to identify the flavour pattern of those interactions. We consider the case in which a strong dynamics lies behind a light Higgs, and explore the implications within the Minimal Flavour Violation ansatz. The dominant effects on flavour-changing Higgs-fermion couplings stem in this context from operators with mass dimension $\le 5$, and we analyze all relevant chiral operators up to that order, including loop-corrections induced by 4-dimensional ones. Bounds on the operator coefficients are derived from a plethora of low-energy flavour transitions, providing a guideline on which flavour-changing Higgs interactions may be open to experimental scrutiny. In particular, the coefficient of a genuinely CP-odd operator is only softly constrained and therefore its impact is potentially interesting.
\end{abstract}
\end{titlepage}
\setcounter{footnote}{0}

\pdfbookmark[1]{Table of Contents}{tableofcontents}
\tableofcontents

%
%
\section{Introduction}

A new resonance at the Electroweak (EW) scale has been established at LHC. Both ATLAS and CMS collaborations 
have recently presented \cite{:2012gk,:2012gu} the discovery of an excess of events above the expected 
Standard Model (SM) background with a local significance of $5\sigma$ consistent with the hypothesis of 
the SM scalar boson\cite{Englert:1964et, Higgs:1964ia,Higgs:1964pj} (so-called ``Higgs boson'' for short) 
with mass around 125 GeV. 

This resonance is, at the moment, compatible with the SM Higgs interpretation, even if the rate in the 
di-photon channel, slightly above SM expectations, leaves still open the possibility of non-standard 
effects, and furthermore a $\sim2\sigma$ tension persists between the predictions and measurement of 
the rate  $R_b^0$ and the forward-backward asymmetry $A_{FB}^{0,b}$, in b-quark production from $e^+-e^-$ 
collisions~\cite{Eberhardt:2012gv,Baak:2012kk}.

There are essentially two main frameworks that have been proposed over the  last decades in order to explain 
the EW symmetry breaking sector. The first possibility is that the Higgs is a fundamental particle, transforming 
linearly (as a doublet in the standard minimal picture) under the gauge symmetry group $SU(2)_L \times U(1)_Y$. 
This line of thought suggests, due to the appearance of the hierarchy problem, to invoke new 
physics (NP) around the TeV scale in order to definitively stabilize the Higgs (and the EW) mass scale. 
The MSSM and its variations are the best explored options of that kind, and a plethora of SUSY partners 
should populate the scale unveiled by LHC experiments, unless awkward fine-tuning effects take place.
 
An interesting alternative is that the Higgs dynamics is non-perturbative and associated to a strong 
interacting force with scale $\Lambda_s$, and the gauge symmetry in the scalar sector is non-linearly 
realized. In the original ``Technicolor" formulation \cite{Susskind:1978ms,Dimopoulos:1979es,Dimopoulos:1981xc}, no 
physical Higgs particle appears in the low-energy spectrum and only the three would-be-Goldstone bosons 
responsible for the weak gauge boson masses are retained. The characteristic scale $f$ associated to the 
Goldstone bosons was identified with the electroweak scale $f=v\equiv246$ GeV, defined from the $W$ mass 
$M_W=g v/2$, and respecting $f \ge \Lambda_s/4 \pi$~\cite{Manohar:1983md}. The smoking gun signature of this Technicolor ansatz is the appearance of several vector 
and fermion resonances at the TeV scale. The discovery of a light Higgs candidate has recently focused 
the attention on an interesting variant: to consider still a strong dynamics behind the electroweak scalar sector 
but resulting -in addition- in a composite (instead of elementary) and light Higgs particle. In this 
scenario, proposed long ago~\cite{Kaplan:1983fs,Kaplan:1983sm,Banks:1984gj,
Georgi:1984ef,Georgi:1984af,Dugan:1984hq}, the Higgs itself would be one of the Goldstone bosons associated 
with the strong dynamics at the scale $\Lambda_s$, while its mass would result from some explicit breaking 
of the underlying strong dynamics. It was suggested that this breaking may be caused by the weak gauge 
interactions or alternatively by  non-renormalizable couplings. These ideas have been revived in recent 
years and are opportune given the recent experimental data (see for example Ref.~\cite{Contino:2010rs} 
for a recent review on the subject).  In this class of scenarios, $f$ may lie around the TeV regime, while $v$ is linked 
to the electroweak symmetry breaking process and is not identified with $f$, $v\le f$. The degree of 
non-linearity is then quantified by a new parameter,
\beq
\xi\equiv \frac{v^2}{f^2},
\label{xi}
\eeq
and, for instance, $f\sim v$ characterizes the extreme non-linear constructions, while $f\gg v$ is typical of 
scenarios which mimic the linear regime. As a result, for non-negligible $\xi$ there may be corrections 
to the size of the SM couplings observable at low energies due to NP contributions. 

The question we address in this paper is the flavour structure of the NP operator coefficients, 
when a strong dynamics is assumed at the scale $\Lambda_s$ and in the presence of a light Higgs particle. 
In particular, dangerous NP contributions to flavour-changing observables could arise. Indeed, the core 
of the flavour problem in NP theories consists in explaining the high level of suppression that must 
be encoded in most of the theories beyond the SM  in order to pass flavour changing neutral current (FCNC) 
observability tests. Minimal Flavour Violation (MFV)~\cite{Chivukula:1987py,Hall:1990ac,D'Ambrosio:2002ex} 
emerged in the last years as one of the most promising working frameworks and it will be used in this work. 

Following the MFV ansatz, flavour in the SM {\it and beyond} is described  at low-energies uniquely in terms 
of the known fermion mass hierarchies and mixings. An outcome of the MFV ansatz is that the energy scale 
of the NP may be as low as few TeV in several distinct contexts \cite{Lalak:2010bk,Fitzpatrick:2007sa,
Grinstein:2010ve,Buras:2011wi}, while in general it should be larger than hundreds of TeV~\cite{Isidori:2010kg}. MFV has been codified as a general framework built upon the flavour symmetry of the kinetic terms~
\cite{Cirigliano:2005ck,Davidson:2006bd,Kagan:2009bn,Gavela:2009cd,Feldmann:2009dc,Alonso:2011yg,Alonso:2011jd,Alonso:2012fy}. For quarks, the flavour group
\beq
G_f=SU(3)_{Q_L}\times SU(3)_{U_R}\times SU(3)_{D_R}
\eeq
defines the non-abelian transformation properties of the $SU(2)_L$ doublet $Q_L$ and singlets $U_R$ and $D_R$,
\beq
Q_L\sim(3,1,1)\,,\qquad\qquad
U_R\sim(1,3,1)\,,\qquad\qquad
D_R\sim(1,1,3)\,.
\eeq
To introduce the Yukawa Lagrangian without explicitly breaking $G_f$, the Yukawa matrices for up ($Y_U$) 
and down ($Y_D$) quarks can be promoted to be spurion fields transforming under the flavour symmetry,
\beq
Y_U\sim(3,\bar3,1)\,,\qquad\qquad
Y_D\sim(3,1,\bar3)\,.
\label{YTransf}
\eeq
The quark masses and mixings are correctly 
reproduced  
once these spurion fields get background values as
\beq
Y_U=V^\dag\,\cy_U\,,\qquad\qquad
Y_D=\cy_D\,,
\eeq
where $\cy_{U,D}$ are diagonal matrices whose elements are the Yukawa eigenvalues, and $V$ a unitary matrix 
that in good approximation coincides with the CKM matrix. These background values break the flavour group 
$G_f$, providing contributions to FCNC observables suppressed by specific combinations of quark mass 
hierarchies and mixing angles. In Ref.~\cite{D'Ambrosio:2002ex}, the complete basis of gauge-invariant 
6-dimensional FCNC operators has been constructed for the case of a linearly realized SM Higgs sector, 
 in terms of the SM fields and the $Y_U$ and $Y_D$ spurions. Operators of dimension $d>6$ are usually 
neglected due to the additional suppression in terms of the cut-off scale.

The MFV ansatz in the presence on a strong interacting dynamics has been introduced in Ref.~\cite{Alonso:2012jc}, 
where the list of relevant $d=4$ flavour-changing operators was identified, in the limit in which the Higgs 
degree of freedom is integrated out. In the non-linear regime a chiral expansion is pertinent, and this results in a different set of operators at leading order than in the case of the linear regime, as the leading operators in the linear and non-linear expansion do not match one-to-one (see for instance the discussion in Ref.~\cite{Alonso:2012px}).  
The promotion of the Yukawa matrices to  spurions follows the same 
lines as in the linear regime, though. Indeed, when the SM quarks $\Psi_{L,R}$ couple bi-linearly to the strong sector\footnote{A more 
complicated case in which the link among the proto-Yukawa interactions and the spurions $Y_{U,D}$ is less 
direct happens in the context of the partial compositeness \cite{Kaplan:1991dc}. In this 
case, quarks couple to the strong sector linearly and therefore two Yukawa couplings, $Y_{\Psi_L}$ and 
$Y_{\Psi_R}$, for each flavour sector appear. Spurions are then identified with only one of these 
proto-Yukawa couplings for each flavour sector, with the other assumed flavour diagonal~\cite{Redi:2011zi}. 
We will not consider here this possibility.}
\beq
\bar\Psi_{L} \,Y_\Psi \Psi_{R}\,\Theta_s\,,
\eeq
with $\Theta_s$ a flavour blind operator in the strong sector, then all flavour information is encoded 
in $Y_\Psi$, that, in order to preserve the flavour group $G_f$, must transform as in Eq.~(\ref{YTransf}). 
Once the spurions have been defined as the only sources of flavour violation (in the SM and beyond), it is 
possible to build the tower of FCNC operators, invariant under both the gauge and the flavour symmetries. 
It is customary to define:
\beq
\lambda_F\equiv Y_U\,Y_U^\dag+Y_D\,Y_D^\dag=V^\dag\cy_U^2 V+\cy_D^2,
\eeq
which transforms as a $(8,1,1)$ under $G_f$. The only relevant non-diagonal entries are all proportional 
to the top Yukawa coupling, $\left(\lambda_F\right)_{ij}\approx y_t^2\,V^*_{ti}\,V_{tj}$, for $i\neq j$.

Within the spirit of MFV, the flavour structure of all Yukawa terms will be dictated only by its fermion composition; in consequence, the resulting fermion-$h$ couplings get diagonalized together with the fermion mass matrix  diagonalization. In other words, flavour-changing couplings require operators of (at least) dimension 5. This property will also apply to the  non-linear analysis below.

 In this work we construct the tower of  $d\le5$ $h$-fermion flavour-changing operators  for a generic strong interacting light Higgs scenario. Which operator basis is chosen in an effective Lagrangian approach is an issue relevant to get the best bounds from a given set of observables, and a convenient basis will be used when analyzing flavour, distinct from that applied in  Refs.~\cite{Giudice:2007fh,Contino:2010mh,Azatov:2012bz,Alonso:2012px} to analyze the Higgs-gauge sector.  A consistent approach requires to   
 revisit as well the $d=4$ flavour-changing operators presented in 
Ref.~\cite{Alonso:2012jc},  by introducing the possibility of a light scalar Higgs, and to consider in addition their main loop-induced effects.  
In the theoretical discussion we will reconsider the interesting exercise performed in Refs.~\cite{Grinstein:2007iv,Contino:2010mh,Azatov:2012bz} to reach the non-linear regime from the linear one~\cite{Giudice:2007fh} in the presence of a light Higgs. 
 We will also perform the phenomenological analysis of the strength of the NP fermionic couplings, focusing  on the -often stringent- bounds on the operator coefficients that follow from present low-energy measurements on the Higgsless component of the couplings. This will provide a guideline on which type of flavoured Higgs couplings may be at reach at the LHC.

The structure of the paper is the following. Sect.~\ref{Framework} describes the framework  and it is mainly devoted to the relation between the linear and non-linear realizations of the electroweak symmetry breaking 
mechanism with a light scalar Higgs particle. Sect.~\ref{OperatorsSection} 
 identifies the $d=4$  and $d=5$  flavour-changing couplings. The main phenomenological impact of both $d=4$ and $d=5$ operators is presented in in Sect.~\ref{PhemSection}. Finally, we conclude in Sect.~\ref{Conclusions}. Technical details on the relation with the SILH Lagrangian~\cite{Giudice:2007fh} can be found in App.~\ref{AppA}; the gauge field equations of motion in the presence of flavour-changing contributions are described in App.~\ref{AppB}; the identification of the $d\ge6$  operators of the linear expansion which correspond to $d=5$ operators of the non-linear one can be found in App.~\ref{AppC}; the relation between the $d=5$ operator coefficients and the corresponding coefficients in the unitary basis is detailed in App.~\ref{AppD}.

%
%

\section{The framework}
\label{Framework}

By ``Higgs'' we mean here a particle that, at some level, participates in the EW symmetry breaking 
mechanism, which requires an $SU(2)$ doublet structure. When building up the hybrid situation in which a non-linear dynamics is assumed but the Higgs is light two strategies are possible: to go from a linear expansion towards a non-linear one, or conversely to start from the non-linear realization of the Goldstone boson mechanism and modify it to account for a light Higgs.  In general, four (related) scales may be relevant, $\Lambda_s$, $f$, $\mean{h}$ and $v$:
\begin{itemize}
\item[i)] $\Lambda_s$ is the energy scale of the strong dynamics and  the typical size of the mass of the strong scalar and fermionic resonances (in the context of QCD, it corresponds to $\Lambda_{\chi SB}$, the scale of the chiral symmetry breaking \cite{Manohar:1983md}).
\item[ii)] $f$ is the characteristic scale associated to the Goldstone bosons that give mass to the gauge bosons and respects $\Lambda_s\leq4\pi f$ (in the context of QCD, it corresponds to the pion coupling constant  $f_\pi$).
\item[iii)] $\mean{h}$ refers to the order parameter of EW symmetry breaking, around which the physical scalar $h$ oscillates.\item[iiii)]  $v$ denotes the EW scale, defined through $M_W= gv/2$. In a general model $\mean{h} \ne v$ and this leads to an $\mean{h}$ dependence in the low-energy Lagrangian through a generic functional form $\cF (h + \mean{h})$.
\end{itemize}

In non-linear realizations such as Technicolor-like models, it may happen that $\mean{h}=v=f$. In the setup considered here with a light $h$ they do not need to coincide, though, although typically a relation links $v$, $\mean{h}$ and $f$.  Thus, a total of three scales will be useful in the analysis, for instance $\Lambda_s$, $f$ and $v$. Without referring to a specific model, one can attempt to describe  the NP impact at low energies resorting to an effective Lagrangian approach, with operators made out of SM fields and invariant under the SM gauge symmetry. The transformation properties of the three longitudinal degrees of freedom of the weak gauge bosons can still be described at low-energy\footnote{Notice that in this low-energy expression for $\UH(x)$, the scale associated to the eaten GBs is $v$ and not $f$. Technically, the scale $v$ appears through a redefinition of the GB fields so as to have canonically normalized kinetic terms.} by a dimensionless unitary matrix transforming as a representation of the global symmetry group:
\beq
\UH(x)=e^{i\sigma_a \pi^a(x)/v}\, , \qquad \qquad  \UH(x) \rightarrow L\, \UH(x) R^\dagger\, ,
\label{UH}
\eeq
with $L,R$ denoting respectively the global transformations $SU(2)_{L,R}$. The adimensionality of $\UH(x)$ is the key to understand that the dimension of the leading  low-energy operators describing the dynamics of the scalar sector differs for a non-linear Higgs sector~\cite{Appelquist:1980vg,Longhitano:1980iz,Longhitano:1980tm,Feruglio:1992wf,Appelquist:1993ka} and a purely linear regime, as insertions of $\UH(x)$ do not exhibit a scale suppression. 

It is becoming customary to parametrize the Lagrangian describing a light dynamical Higgs particle $h$  by means of the following ansatz~\cite{Contino:2010mh,Azatov:2012bz} (see also Ref.~\cite{Grinstein:2007iv}):
\beq
\begin{aligned}
\LL_h=&\phantom{+\,\,\,}\frac{1}{2} (\partial_\mu h) (\partial^\mu h) \,\left(1+c_H\,\xi\,\cF_H(h)\right)- V(h)+\\
&-\frac{v^2}{4}\Tr\left[\VL^\mu \VL_\mu\right] \,\cF_C(h)+c_T\frac{v^2}{4}\,\xi\, \Tr\left[\TL\VL^\mu\right]\Tr\left[\TL\VL_\mu\right] \cF_T(h)+\\
& - \left\{\frac{v}{\sqrt{2}}\bar{Q}_L\UH(x)\,\cY\,\diag(\cF^U_Y(h),\cF^D_Y(h))\, Q_R+\mbox{h.c.} \right\} +\ldots\,.
\end{aligned}
\label{Lagh}
\eeq
where dots stand for higher order terms, and \mbox{$\VL_\mu\equiv \left(\DL_\mu\UH\right)\UH^\dagger$} ($\TL\equiv\UH\sigma_3\UH^\dag$) is the vector (scalar) chiral field transforming in the adjoint of the gauge group  $SU(2)_L$. The covariant derivative reads 
\beq
\DL_\mu \UH(x) \equiv \derp_\mu \UH(x) +\dfrac{ig}{2}\,W_{\mu}^a(x)\,\sigma_a\,\UH(x) - 
                      \dfrac{ig'}{2} \,B_\mu(x) \, \UH(x)\,\sigma_3 \, ,
\label{cDUH}
\eeq
with $W^a_\mu$ ($B_\mu$) denoting the $SU(2)_L$ ($U(1)_Y)$ gauge bosons and $g$ ($g'$) the corresponding gauge coupling.  
In these equations, $V(h)$ denotes  the effective scalar potential describing the breaking of the electroweak symmetry,  the first term in Eq.~(\ref{Lagh}) includes the Higgs kinetic term, while the second line describes the $W$ and $Z$ masses and their interactions with $h$ as well as the usual custodial symmetry breaking term. Finally, restricting our considerations to the quark sector, the third line accounts for the Yukawa-like interactions between $h$ and the SM quarks, grouped in doublets of the global symmetry $Q_{L,R}$, with
\beq
\cY\equiv\diag\left(Y_U,\,Y_D\right)\,,
\nn
\eeq
$Y_U$ and $Y_D$ being the usual Yukawa matrices. The parameters $c_H$ and $c_T$ are model dependent operator coefficients.

The functions $\cF_i(h)$ in Eq.~(\ref{Lagh}), as well as other $\cF(h)$ functions defined below, encode the generic dependence on the light $h$ particle. Each $\cF(h)$ function can be  expanded in powers of $\xi$,  $\cF(h)= g_0(h,v) + \xi g_1(h,v) + \xi^2 g_2(h,v) + \ldots$, where $g(h,v)$ are model-dependent functions of $h$. We will not need to enter in their precise dependence in this work; a discussion can be found in Ref.~\cite{Alonso:2012px} and references therein. We just mention here that in previous literature~\cite{Contino:2010mh,Azatov:2012bz} the functional dependence of some of those 
functions has been expressed as a power series in $h/v$: 
\beq
\begin{aligned}
\cF_C(h)&= \left(1+2a\,\frac{h}{v}+b\,\frac{h^2}{v^2} + \dots \right)\,,\\
\cF^{U,D}_Y(h)&=\left(1+c^{U,D}\,\dfrac{h}{v}+\ldots\right)\,.
\end{aligned}
\nn
\eeq
The constants $a$, $b$ and $c$ are model-dependent parameters and encode the dependence on $\xi$. 
The $a$ and $c_T$ parameters are constrained from electroweak precision tests: in particular 
$0.7\lesssim a\lesssim1.2$ \cite{Azatov:2012qz} and $-1.7\times10^{-3}<c_T\,\xi<1.9\times10^{-3}$ 
\cite{Giudice:2007fh} at $95\%$ CL.

The Lagrangian discussed above can be very useful to describe an extended class of ``Higgs'' models, ranging from the SM  scenario (for $\mean{h}=v$, $a=b=c^{U,D}=1$ and neglecting  higher order terms in $h$), to the Technicolor-like ansatz (for $f\sim v$ and omitting all terms in $h$) and intermediate situations with a light scalar $h$ (in general for $f\ne v$) as in composite/holographic Higgs models \cite{Dimopoulos:1981xc,Kaplan:1983fs,Kaplan:1983sm,Georgi:1984ef,Banks:1984gj,Georgi:1984af,Dugan:1984hq,Agashe:2004rs,Contino:2006qr,Gripaios:2009pe} up to dilaton-like scalar frameworks \cite{Halyo:1991pc,
Goldberger:2007zk,Vecchi:2010gj,Campbell:2011iw,Matsuzaki:2012mk,Chacko:2012vm,Bellazzini:2012vz}. Note that, although electroweak radiative corrections severely constraint Technicolor-like scenarios, in concrete models values of $v/f$ as large as $v/f\sim 0.4-0.6$ are still allowed at the price of acceptable $10\%$ fine-tunings \cite{Contino:2010rs,Panico:2012uw}. As a result, the study of higher dimension operators is strongly motivated, especially as the limits on $\xi$ are quite model-dependent: in the effective Lagrangian approach $\xi$ will be left free $0<\xi<1$ while the constraints on custodial breaking effects will be translated into limits on the operator coefficients. For the case of  pure gauge and $h$-gauge couplings,  some of the couplings have been explicitly explored in Refs~\cite{Giudice:2007fh,Contino:2010mh,Azatov:2012bz} and a complete basis of independent operators up to dimension five has been provided in Ref.~\cite{Alonso:2012px}. 

The $\xi$ parameter in Eq.~(\ref{xi}) defines the degree of non-linearity of a specific model and in particular $\xi \rightarrow 0 $ refers to the linear regime, while $\xi \rightarrow 1 $ to the non-linear one. For $\xi \ll 1$ the hierarchy between operators mimics that in the linear expansion, where the operators are written in terms of the Higgs doublets $H$: couplings with higher number of (physical) Higgs legs are suppressed compared to the SM renormalizable ones, through powers of the high NP scale or, in other words, of $\xi$~\cite{Manohar:1983md}. The power of $\xi$ keeps then track of the $h$-dependence of the $d>4$ operators, where the insertions of $h$ enter only through powers of $(\mean{h}+h)/f\simeq\xi^{1/2}(v+h)/v$, and of $\partial_\mu h/f^2$  (see Ref.~\cite{Alonso:2012px}). In the $\xi \ll 1$ limit, the $\cF_i(h)$ functions, appearing in Eq.~(\ref{Lagh}) and in the following, would inherit the same universal behaviour in powers of $(1+h/v)$: at order $\xi$, that is, for couplings that would correspond to $d=6$ operators of the linear expansion, it follows that 
\beq
\cF_i(h)=F(h)\equiv\left(1+\frac{h}{v}\right)^2\,,
\label{Fs}
\eeq
An obvious extrapolation applies to the case of couplings weighted by higher powers of $\xi$, that is with $d>6$.

When $\xi \approx 1$ the $\xi$ dependence does not entail a suppression of operators compared to the renormalizable SM operators and the chiral expansion should instead be adopted, although it should be clarified at which level the effective expansion on $h/f$ should stop. Below, the $\cF(h)$ functions will be considered completely general polynomial of $\mean{h}$ and $h$ (in particular not of derivatives of $h$) and, when using equations of motion and integration by parts to relate operators, they would be assumed to be redefined when convenient, much as one customarily redefines the constant operator coefficients. 

To analyze the passage from the linear to the non-linear regime, it is an interesting exercise to explore the transition from a $SU(2)_L\times U(1)_Y$ invariant effective Lagrangian in the linear realization of the EW symmetry breaking mechanism to an effective chiral Lagrangian. For instance, in the so-called SILH framework, operators may be written in either the linear \cite{Giudice:2007fh} (i.e. using the $H(x)$ doublet) or the non-linear \cite{Contino:2010mh,Azatov:2012bz} (i.e. using the $\UH$ matrix and a  scalar field $h$) formalism.  We have revisited this procedure in App.~\ref{AppA}.

\section{The flavour sector}
\label{OperatorsSection}

The choice of operator basis most suitable when analyzing fermionic couplings is in general one in which fermionic fields participate in the operators. 

The flavour-changing  sector has not been explicitly taken into consideration in previous analysis of the effective Lagrangian for a  strong interacting light Higgs. Flavour-changing terms do appear in the equations of motion for the gauge field strengths in the presence of the effective operators, and the explicit expressions can be found in Appendix \ref{AppB}. However, to include them explicitly would translate into corrections to flavoured observables that are quadratic in the effective operator coefficients $a_i$, and more precisely of the type $\cO(a_{FC} \times a_{GH})$, where $a_{FC}$ ($a_{GH}$) represent the generic flavour-changing (gauge-$h$) coefficients. Given that $a_{GH}$ are severely constrained by EW data  (barring extreme fine-tunings), those quadratic corrections can be disregarded in the rest of the analysis and it is enough to consider the SM equations of motion:
\begin{align}
\left(D^\mu\,W_{\mu\nu}\right)_j &= i\,\dfrac{g}{4}\,v^2\,\Tr[\VL_\nu\,\sigma_j]\,\cF_C(h) \,+\,
  \dfrac{g}{2}\,\bar Q_L\,\gamma_\nu\,\sigma_j\,Q_L 
\label{triplet} 
\,,\\
\derp^\mu\,B_{\mu\nu}&= -\,i\,\dfrac{g'}{4}\,v^2\,\Tr[\TL\,\VL_\nu]\,\cF_C(h)\,+\, g'\,\bar Q_L\,\gamma_\nu\,
  {\mathbf h}_L \,Q_L \,+\, g'\,\bar Q_R\,\gamma_\nu\,{\mathbf h}_R \,Q_R\,,
\label{singlet}
\end{align}
with ${\mathbf h}_{L,R}$ the left and right hypercharges in the $2\times 2$ matrix notation.
In resume, the analysis of the flavour-changing sector can be considered ``independent" of that for the gauge-$h$ and flavour-conserving sectors.

\boldmath
\subsection{From $d=4$ non-linear operators}
\unboldmath

With the aid of the (Goldstone) chiral fields $\TL$ and $\VL_\mu$ it is only possible to write $d=4$ 
fermionic operators involving two right-handed (RH) or two left-handed (LH) fields. In the MFV framework under consideration, only operators 
built with two LH fermions can induce flavour-changing effects at leading order in the spurion expansion. 
Consequently,  terms with two RH fermions will not be considered in what follows.

A total of four independent $d=4$ chiral operators containing LH fermion fields can be constructed 
\cite{Appelquist:1984rr,Cvetic:1988ey,Espriu:2000fq,Alonso:2012jc}, namely:
\beq
\begin{aligned}
&\mathcal{O}_{1}=\frac{i}{2}\,\bar{Q}_L\,\lambda_{F}\,\gamma^{\mu}\,\left\{\TL,\VL_{\mu} \right\}\, Q_L\,, 
&\qquad & \mathcal{O}_{2}=i\,\bar{Q}_L\,\lambda_{F}\,\gamma^{\mu}\,\VL_{\mu}\,Q_L\,, \\
&\mathcal{O}_{3}=i\,\bar{Q}_L\,\lambda_{F}\,\gamma^{\mu}\,\TL\,\VL_{\mu}\,\TL\, Q_L\,, & \qquad &
\mathcal{O}_{4}=\frac{1}{2}\,\bar{Q}_L\,\lambda_{F}\,\gamma^{\mu}\,\left[\TL,\VL_{\mu}\right]\,Q_L\,.
\end{aligned}
\label{d4Operators}
\eeq
Out of these $\cO_1-\cO_3$ are CP-even while $\cO_4$ is intrinsically CP-odd \cite{Alonso:2012jc}.

Following the discussion  in Sec.~\ref{Framework}, it is pertinent to extend the definition of these chiral 
couplings in order to include the possibility of a light scalar degree of freedom (related to the EW 
symmetry breaking), through the ansatz:
\beq
\LL_{\chi=4}^f=\xi \sum_{i=1,2, 3} \hat{a}_i\,\cO_i(h)+\xi^2  \hat{a}_4\,\cO_4(h)
\label{d4OperatorsH}
\eeq  
where a redefinition by powers of $\xi$ of the operators coefficients defined in Ref.~\cite{Alonso:2012jc} has been implemented,
$a_i\equiv \xi \, \hat{a_i}$ for $i=1,2,3 $, while  $a_4\equiv \xi^2 \, \hat{a_4}$. Furthermore  
\beq
\cO_i(h)\equiv \cO_i\,\cF_i(h)\,,
\label{defOh}
\eeq 
where again the functions $\cF_i(h)$ contain  the dependence on $(h + \mean{h})$. In the present work --restrained to effective couplings of total dimension $d\le5$-- only terms linear in $h$ should be retained in Eq.~(\ref{d4OperatorsH}); for the same reason it is neither  pertinent to consider couplings containing $\partial_\mu h$ (that is, derivatives of $\cF(h)$). 
For $\xi \ll 1$, the functions $\cF_i(h)$ collapse into combinations of $F(h)$ as defined in Eq.~(\ref{Fs}) for the linear regime: 
\beq
\begin{aligned}
&\cO_1(h)\equiv \cO_1\, F(h)\left(1 + \alpha_1\, \xi \, F(h)\right)\,, &\qquad &
 \cO_2(h)\equiv \cO_2\, F(h)\left(1 + \alpha_2\, \xi \, F(h)\right)\,, \\
&\cO_3(h)\equiv \cO_3\,F(h)\left(1 + \alpha_3\, \xi \, F(h)\right)\,, &\qquad & 
 \cO_4(h)\equiv \cO_4\,F^2(h)\,.
\end{aligned}
\label{d4OperatorsH-linear}
\eeq
The powers of $\xi$ in Eqs.~(\ref{d4OperatorsH}) and (\ref{d4OperatorsH-linear}) facilitate the identification of the lowest dimension
at which a ``sibling'' operator appears in the linear regime. By sibling we mean an operator written in terms of $H$, that includes  the couplings $\cO_{1-4}$. For instance, the lowest-dimension siblings of $\cO_1$ and $\cO_2$ arise at $d=6$, while that of $\cO_4$ appears at $d=8$~\cite{Alonso:2012jc}. The case of $\cO_3$ is special: indeed, it corresponds to a combination of a $d=6$ and a $d=8$ operators of the linear expansion. The parametrization in Eq.~(\ref{d4OperatorsH-linear}) reflects this correspondence, where for all the operators the contributions from siblings up to $d=8$ have been accounted for (further contributions will arise considering higher-dimension siblings).

For  $\xi\ll 1$ it is consistent to retain just the terms linear in $\xi$ and neglect the contributions from  $\cO_4(h)$, while it can be shown~\cite{Alonso:2012jc} that $\cO_3(h)$ coincides with $-\cO_2(h)$ and finally only two linearly-independent flavoured operators remain (e.g. $\cO_1(h)$ and $\cO_2$(h)), as previously studied in the literature. On the contrary, in the $\xi \sim 1$ limit all four operators are on the same footing, higher order terms in $\xi$ may contribute,  and one 
recognizes the need of a QCD-like resummation. In particular any chiral operator is made up by an infinite 
combination of linear ones, an effect represented by the  generic $\cF_i(h)$ functions, which admit  in general an expansion in powers of $\xi$ as discussed previously.

In Ref.~\cite{Alonso:2012jc} we set limits on the coefficients of 
the operators  $\cO_1-\cO_4$ from the analysis of $\Delta F=1$ and $\Delta F=2$ observables. The inclusion 
of a light scalar $h$ does not modify the bounds obtained there for the overall coefficients. In fact, the overall operator coefficients  in Eq.~(\ref{d4OperatorsH}) may differ form their Higgsless counterparts in Eqs.~(\ref{d4Operators}) only through a (negligible) loop contribution.

With the inclusion of the light $h$ field, the low-energy effective flavour Lagrangian induced by the SM and the $\cO_1(h)-\cO_4(h)$ operators in Eq.~(\ref{defOh}) reads,  in the unitary gauge (i.e. $\UH(x) = \unity$) and up to $d=5$ couplings, 
\beq
\begin{split}
\LL^f_{\chi=4} =&
 - \frac{g}{\sqrt{2}}\left[ W^+_\mu\bar{U}_L \gamma^\mu [a_W(1+\beta_W\,h/v)+
 ia_{CP}(1+\beta_{CP}\,h/v)] \left(\mathbf{y}_U^2 V +
 V\mathbf{y}_D^2\right) D_L + h.c. \right]+ \\
& -\frac{g}{2\cos\theta_W}Z_\mu\left[a^{u}_Z\bar{U}_{L}\gamma^\mu \left(\mathbf{y}_U^2+
V\mathbf{y}_D^2 V^\dagger\right) U_L\,(1+ \beta^{u}_Z\,h/v)\right.\\
&\qquad \qquad\qquad\left.+ a_Z^d\bar{D}_{L}\gamma^\mu \left(\mathbf{y}_D^2+
V^\dagger\mathbf{y}_U^2 V\right) D_L\,(1+ \beta^{u}_Z\,h/v)\right]\,,
\end{split}
\label{DevSM}
\eeq
where 
\beq
\begin{aligned}
a_Z^u&\equiv a_1+a_2+a_3\,,&\qquad\qquad
a_Z^d&\equiv a_1-a_2-a_3\,,\\
a_W&\equiv a_2-a_3\,, &\qquad\qquad
a_{CP}&\equiv - a_4\,.
\end{aligned}
\label{ChangeBasisZW}
\eeq
The arbitrary coefficients $\beta_i$ in Eq.~(\ref{DevSM}) follow a similar rearrangement to that for $a_i$ in Eq.~(\ref{ChangeBasisZW}), once the $\cF(h)$ functions are expanded to first order in $h$, $\cF_i(h)\sim (1 + \beta_i\,h+...)$; in  general each $\beta_i$ may receive contributions from all orders in $\xi$ for large $\xi$.

All limits obtained in Ref.~\cite{Alonso:2012jc} for the values of $a_Z^d$, $a_W$ and $a_{CP}$ resulted from tree-level contributions to observables. It is interesting -and necessary when considering $d=5$ effective couplings- to analyze as well the possible bounds on the $a_i$ coefficients from their contribution (still disregarding the $h$ insertions) at loop-level to radiative processes, such as $b\to s\gamma$ decay.  Indeed, the modification of the CKM matrix has a non-negligible impact in the branching ratio of this observable and its precision on both the experimental determination and the theoretical prediction constrains significantly the $a_W-a_{CP}$ parameter space, as we will show in Sec.~\ref{PhemSection}.

Finally, an important difference with strongly interacting heavy Higgs scenarios is the presence at low-energies of vertices with additional $h$ external legs, as indicated by Eq.~(\ref{DevSM}). This implies interesting phenomenological consequences that will be illustrated later on.

\boldmath
\subsection{From $d=5$ non-linear operators}
\label{subsecd5}
\unboldmath

To our knowledge, no discussion of $d=5$ flavour-changing {\it chiral} operators has  been presented in literature.  They may contribute at tree-level  to relevant flavour-changing observables, as for instance the $b \to s \gamma$ branching ratio. In this subsection  all  $d=5$ flavour-changing chiral operators are identified, while interesting phenomenological consequences will be discussed in Sect.~\ref{PhemSection}.

Gauge invariant $d=5$ operators relevant for flavour must have a bilinear 
structure in the quark fields of the type $\bar Q_L\,(\cdot \cdot \cdot)\,\UH(x)\,Q_R$, where dots stand 
for objects that transform in the trivial or in the adjoint representation of $SU(2)_L$. Besides the 
vector and scalar chiral fields $\VL_\mu$ and $\TL$, they can contain either the rank-2 antisymmetric 
tensor $\tmn$ or the strength tensors $B_{\mu\nu}$, $W_{\mu\nu}$ and $G_{\mu\nu}$. 
According to their Lorentz structure, the resulting independent $d=5$ chiral couplings can be classified in three main groups: 
\begin{description}
\item[i) dipole-type operators:]
\beq
\begin{aligned}
&\mathcal{X}_{1}=  g'\,\bar{Q}_L\,\tmnp\,\UH\,Q_R \,B_{\mu\nu}\,,\qquad\qquad
&&\mathcal{X}_{2}= g' \,\bar{Q}_L\,\tmnp\,\TL\,\UH\,Q_R\,B_{\mu\nu}\,,\\
&\mathcal{X}_{3}=  g \,\bar{Q}_L\,\tmnp\,\sigma_i\UH\,Q_R\,W^i_{\mu\nu}\,,\qquad\qquad
&&\mathcal{X}_{4}= g \,\bar{Q}_L\,\tmnp\,\sigma_i\TL\,\UH\,Q_R\,W^i_{\mu\nu}\,,\\
&\mathcal{X}_{5}=  g_s \,\bar{Q}_L\,\tmnp\,\UH\,Q_R\,G_{\mu\nu}\,,\qquad\qquad
&&\mathcal{X}_{6}= g_s \, \bar{Q}_L\,\tmnp\,\TL\,\UH\,Q_R\,G_{\mu\nu}\,, \\
&\mathcal{X}_{7}=  g \,\bar{Q}_L\,\tmnp\,\TL\,\sigma_i\,\UH\,Q_R\,W^i_{\mu\nu}\,,\qquad\qquad
&&\mathcal{X}_{8}= g \,\bar{Q}_L\,\tmnp\,\TL\,\sigma_i\TL\,\UH\,Q_R\,W^i_{\mu\nu}\,;
\end{aligned}
\label{OpGroup3}
\eeq
\item[ii) operators containing the rank-2 antisymmetric tensor $\tmnp$:]
\beq
\begin{aligned}
&\mathcal{X}_{9\phantom{0}}=\bar{Q}_L\,\tmnp\,[\VL_\mu,\VL_\nu]\,\UH\,Q_R\,,\qquad
&&\mathcal{X}_{10}=\bar{Q}_L\,\tmnp\,[\VL_\mu,\VL_\nu]\,\TL\,\UH\,Q_R\,,\\
&\mathcal{X}_{11}=\bar{Q}_L\,\tmnp\,[\VL_\mu \,\TL,\VL_\nu \,\TL]\,\UH\,Q_R\,,\qquad
&&\mathcal{X}_{12}=\bar{Q}_L\,\tmnp\,[\VL_\mu \,\TL,\VL_\nu \,\TL]\,\TL\,\UH\,Q_R\,;
\end{aligned}
\label{OpGroup2}
\eeq
\item[iii) other operators containing the chiral vector fields $\VL_\mu$:]
\beq
\begin{aligned}
&\mathcal{X}_{13}=\bar{Q}_L\,\VL_\mu\, \VL^\mu\,\UH \,Q_R\,,\qquad\qquad
&&\mathcal{X}_{14}=\bar{Q}_L\, \VL_\mu\, \VL^\mu\, \TL\,\UH \,Q_R\,,\\
&\mathcal{X}_{15}=\bar{Q}_L\, \VL_\mu\, \TL\, \VL^\mu\,\UH \,Q_R\,,\qquad\qquad 
&&\mathcal{X}_{16}=\bar{Q}_L\, \VL_\mu\,\TL \,\VL^\mu\,\TL\,\UH \,Q_R\,,\\
&\mathcal{X}_{17}=\bar{Q}_L\, \TL \,\VL_\mu \,\TL \,\VL^\mu\,\UH \,Q_R\,,\qquad\qquad 
&&\mathcal{X}_{18}=\bar{Q}_L\, \TL\, \VL_\mu \,\TL \,\VL^\mu \,\TL\,\UH \,Q_R\,. 
\end{aligned}
\label{OpGroup1}
\eeq
\end{description}

A fourth group of operators can be constructed from the antisymmetric rank 2 chiral tensor, 
that transforms in the adjoint of $SU(2)_L$:
\be
\VL_{\mu\nu} \equiv \cD_{\mu}\VL_{\nu}-\cD_{\nu}\VL_{\mu}
             = i\,g\,\WL_{\mu \nu} -i\,\dfrac{g}{2}\,B_{\mu\nu}\,\TL+ \left[\,\VL_\mu, \VL_\nu\right] \,.
\label{chiralRank2}
\ee
However, the second equality in Eq.~(\ref{chiralRank2}) shows that operators including $\VL_{\mu\nu} $ are 
not linearly independent from those listed in Eqs.~(\ref{OpGroup3})-(\ref{OpGroup2}). 

The chiral Lagrangian containing the 18 fermionic flavour-changing $d=5$ operators can thus be written as
\beq
\LL_{\chi=5}^f=\sum_{i=1}^{18}\,b_i\,\dfrac{\cX_i}{\Lambda_s}\,,
\label{OrigLag5}
\eeq
where $\Lambda_s$ is the scale of the strong dynamics and $b_i$ are arbitrary $\cO(1)$ coefficients. It is worth to underline that for the analysis of $d=5$ operators in the non-linear regime, the relevant scale is $\Lambda_s$ and not $f$ as for the analysis in the previous section. Indeed, $f$ is associated to light Higgs insertions, while $\Lambda_s$ refers to the characteristic scale of the strong resonances that, once integrated out, give rise to the operators listed in Eqs.~(\ref{OpGroup3})-(\ref{OpGroup1}).

A redefinition of the coefficients allows to make explicit the connection to their lowest-dimension siblings in the linear expansion: 
\beq
\LL_{\chi=5}^f= \sqrt{\xi}\,\sum_{i=1}^{8}\,\hat{b}_i\,\dfrac{\cX_i}{\Lambda_s}+\xi\sqrt{\xi}\sum_{i=9}^{18}\,\hat{b}_i\,\dfrac{\cX_i}{\Lambda_s}\,. 
\label{OrigLag5-bis}
\eeq
In the limit of small $\xi$, $\cX_{1-6}$ correspond to $d=6$ operators in the linear expansion, while $\cX_{7}$ and $\cX_{8}$ result from combinations of $d=6$ and $d=8$ siblings. Moreover, $\cX_{9-18}$ have linear siblings of $d=8$, but $\cX_{17}$ and $\cX_{18}$ that are combinations of $d=8$ and $d=10$ operators in the linear regime. The complete list of the linear siblings of the chiral $d=5$ operators can be found in Appendix \ref{AppC}.

Because in this work the analysis will be restrained to (at most) $d=5$ couplings, it is not necessary nor pertinent to discuss further  the possible extensions $\mathcal{X}_i\rightarrow \mathcal{X}_i(h)$ that would include the dependence on a light Higgs through generic $\cF_i(h)$ functions.   

The phenomenological impact of these contributions can be best identified through the low-energy Lagrangian 
written in the unitary gauge:
\beq
\delta\LL_{\chi=5} =\delta\LL^u_{\chi=5} +\delta\LL^d_{\chi=5} +\delta\LL^{ud}_{\chi=5} \,,
\eeq
where
\begin{align}
\begin{split}
\delta\LL^d_{\chi=5} =&
\frac{g^2}{4 \cos\theta_W^2}\dfrac{b_Z^d}{\Lambda_s}\bar D_L D_R Z_{\mu}Z^{\mu} + 
\frac{g^2}{2}\dfrac{b_W^d}{\Lambda_s}\bar D_L\,D_R W^+_{\mu}W^{-\mu}+
g^2 \dfrac{c_W^d}{\Lambda_s}\bar D_L\,\tmnp D_R W^+_{\mu}W^{-}_{\nu}+\\
&+ e\dfrac{d_F^d}{\Lambda_s}\bar D_L \,\tmnp D_R F_{\mu\nu} + 
\frac{g}{2 \cos\theta_W}\dfrac{d_Z^d}{\Lambda_s}\bar D_L\,\tmnp D_R Z_{\mu\nu}+
g_s\dfrac{d_G^d}{\Lambda_s}\bar D_L \,\tmnp D_R G_{\mu\nu}+\hc\,,
\end{split}
\label{LELag}\\
\begin{split}
\delta\LL^{ud}_{\chi=5} =&\phantom{+}
\frac{g^2}{2\sqrt{2} \cos\theta_W}\left(\dfrac{b^+_{WZ}}{\Lambda_s}\bar U_L D_R \,W^+_{\mu}Z^{\mu}+
\dfrac{b^-_{WZ}}{\Lambda_s}\bar D_L U_R \,W^-_{\mu}Z^{\mu}\right)+\\
&+\frac{g^2}{2\sqrt{2} \cos\theta_W}\left(\dfrac{c^+_{WZ}}{\Lambda_s}\bar U_L\,\tmnp D_R\,W^+_{\mu}Z_{\nu}+
\dfrac{c^-_{WZ}}{\Lambda_s}\bar D_L\,\tmnp U_R\,W^-_{\mu}Z_{\nu}\right)+\\
&+\frac{g}{\sqrt{2}}\left(\dfrac{d^+_W}{\Lambda_s}\bar U_L \,\tmnp D_R \,W^+_{\mu\nu}+\dfrac{d^-_W}{\Lambda_s}\bar D_L \,\tmnp U_R \,W^-_{\mu\nu}\right)+\hc\,,
\end{split}
\label{LELag2}
\end{align}
and analogously for $\delta\LL^u_{\chi=5}$ as in $\delta\LL^d_{\chi=5}$ interchanging $d\leftrightarrow u$ and $D_{L,R}\leftrightarrow U_{L,R}$. In these equations $W^\pm_{\mu\nu}=\derp_\mu W^\pm_{\nu}-\derp_\nu W^\pm_{\mu}\pm i\,g\, \left(W^3_\mu W^\pm_\nu-W^3_\nu W^\pm_\mu\right)$, while
 the photon and $Z$ field strengths are defined as $F_{\mu\nu}=\derp_\mu A_\nu-\derp_\nu A_\mu$ 
and $Z_{\mu\nu}=\derp_\mu Z_\nu-\derp_\nu Z_\mu$, respectively, and  $W^3_\mu = \cos\theta_W Z_\mu + \sin\theta_W A_\mu$.
 The relations between the coefficients appearing in Eqs.~(\ref{LELag}) and (\ref{LELag2}) and those defined in Eq.~(\ref{OrigLag5}) 
are reported in Appendix \ref{AppD}.

%
%
\section{Phenomenological analysis}
\label{PhemSection}

This section first resumes and updates the bounds existing in the literature~\cite{Alonso:2012jc} on the coefficients of the flavour-changing  $d=4$ chiral expansion,  and then discusses new bounds and other phenomenological considerations  with and without a light Higgs:
\begin{itemize}
\item[-] Loop level impact of fermionic $d=4$ chiral operators ($\cO_1$ to $\cO_4$) on those same radiative decays;
\item[-] Tree-level bounds on the  fermionic $d=5$ chiral operators $\cX_i$, from radiative decays;
\item[-] Light Higgs to fermions couplings, from  operators $\cO_1(h)$ to $\cO_4(h)$.
\end{itemize}

\boldmath
\subsection{$\Delta F=1$ and $\Delta F=2$ observables}
\unboldmath

In Ref.~\cite{Alonso:2012jc}, the constraints on the coefficients of the $d=4$ flavour-changing operators of the non-linear expansion have been analysed. These bounds resulted from $\Delta F=1$ and $\Delta F=2$ observables and apply straightforwardly to non-linear regimes with a light $h$ scalar. Operators $\cO_1$, $\cO_2$ and $\cO_3$ induce tree-level contributions to $\Delta F=1$ processes mediated by the $Z$ boson, as can be seen from the $Z$ couplings in the effective Lagrangian in Eq.~(\ref{DevSM}), and are severely constrained. Due to the MFV structure of the coefficients, sizeable flavour-changing effects may only be expected in the  down quark sectors, with data on $K$ and $B$ transitions providing the strongest constraints on   $a_Z^d$,   
\beq
-0.044<a_Z^d<0.009\qquad\qquad \text{at $95\%$ of C.L.}
\eeq
from $K^+\to\pi^+\bar\nu\nu$, $B\to X_s\ell^+\ell^-$ and $B\to\mu^+\mu^-$ data.

Furthermore, operators $\cO_2$, $\cO_3$ and $\cO_4$ induce   corrections to the fermion-$W$ couplings, and thus to the  the CKM matrix, see Eq.~(\ref{DevSM}). This in turn induces modifications~\cite{Alonso:2012jc} on the strength of meson oscillations (at loop level), on $B^+\to\tau^+\nu$ decay and on the $B$ semileptonic CP-asymmetry, among others; more specifically the following process have been taken into account in Ref.~\cite{Alonso:2012jc}:
\begin{itemize}
\item[-] The CP-violating parameter $\epsilon_K$ of the $K^0-\bar K^0$ system and the mixing-induced CP asymmetries $S_{\psi K_S}$ and $S_{\psi\phi}$ in the decays $B^0_d\to\psi K_S$ and $B^0_s\to\psi\phi$. The corrections induced to $\epsilon_K$ are proportional to $y_t^2$, while those to $S_{\psi K_S}$ and $S_{\psi\phi}$ are proportional to $y_b^2$. Consequently, possible large deviations from the values predicted by the SM  are only allowed  in the $K$ system.

\item[-] The ratio among the meson mass differences in the $B_d$ and $B_s$ systems, $R_{\Delta M_B}\equiv \Delta M_{B_d}/\Delta M_{B_s}$. The NP contributions on the mass differences almost cancel in this ratio and therefore deviations from the SM prediction for this observable are negligible.

\item[-] The ratio among the $B^+\to\tau^+\nu$ branching ratio and the $B_d$ mass difference, $R_{BR/\Delta M}\equiv BR(B^+\to\tau^+\nu)/\Delta M_{B_d}$. This observable is clean from theoretical hadronic uncertainties and the constraints on the NP parameters are therefore potentially strong.
\end{itemize}
Since only small deviations from the SM prediction for $S_{\psi K_S}$ are allowed,  only values close to the exclusive determination for $|V_{ub}|$ are favoured (see Ref.~\cite{Alonso:2012jc} for a complete discussion). Moreover, it is possible to constrain the $|V_{ub}|-\gamma$ parameter space, with $\gamma$ being one of the angles of the unitary triangle, requiring that both $S_{\psi K_S}$ and $R_{\Delta M_B}$ observables are inside the $3\sigma$ experimental determination. 

\begin{figure}[h!]
 \centering
 \subfigure[Correlation plot between $\vep_K$ and $R_{BR/\Delta M}$. $a_W,a_{CP}\in{[}-1,1{]}$, 
            $a_Z^d\in {[}-0.1,0.1{]}$]{
\includegraphics[width=7.6cm]{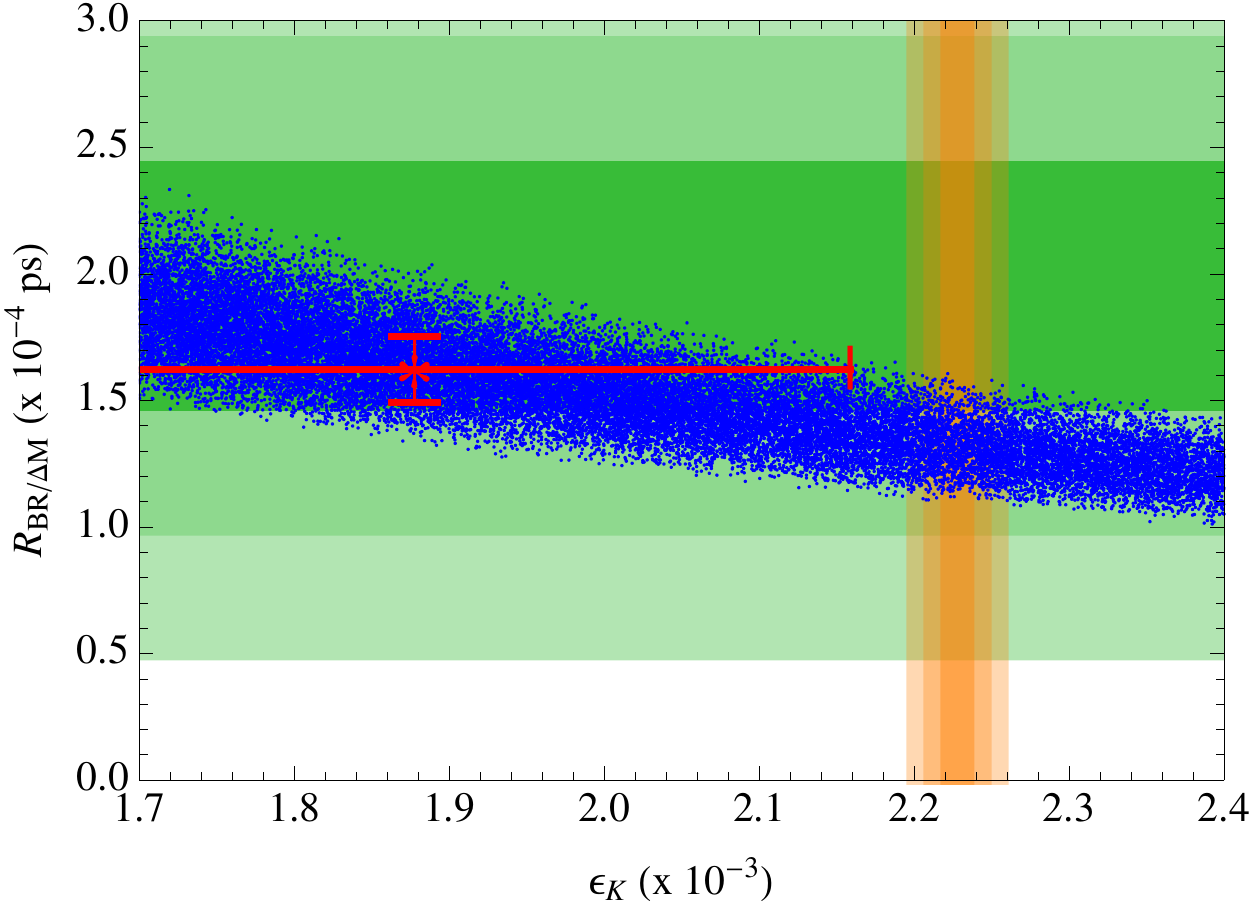}}  
\subfigure[$a_W-a_{CP}$ parameter space for the observables on the left panel inside their $3\sigma$ error ranges and 
            $a_Z^d\in{[}-0.044,0.009{]}$.]{
\includegraphics[width=7.8cm]{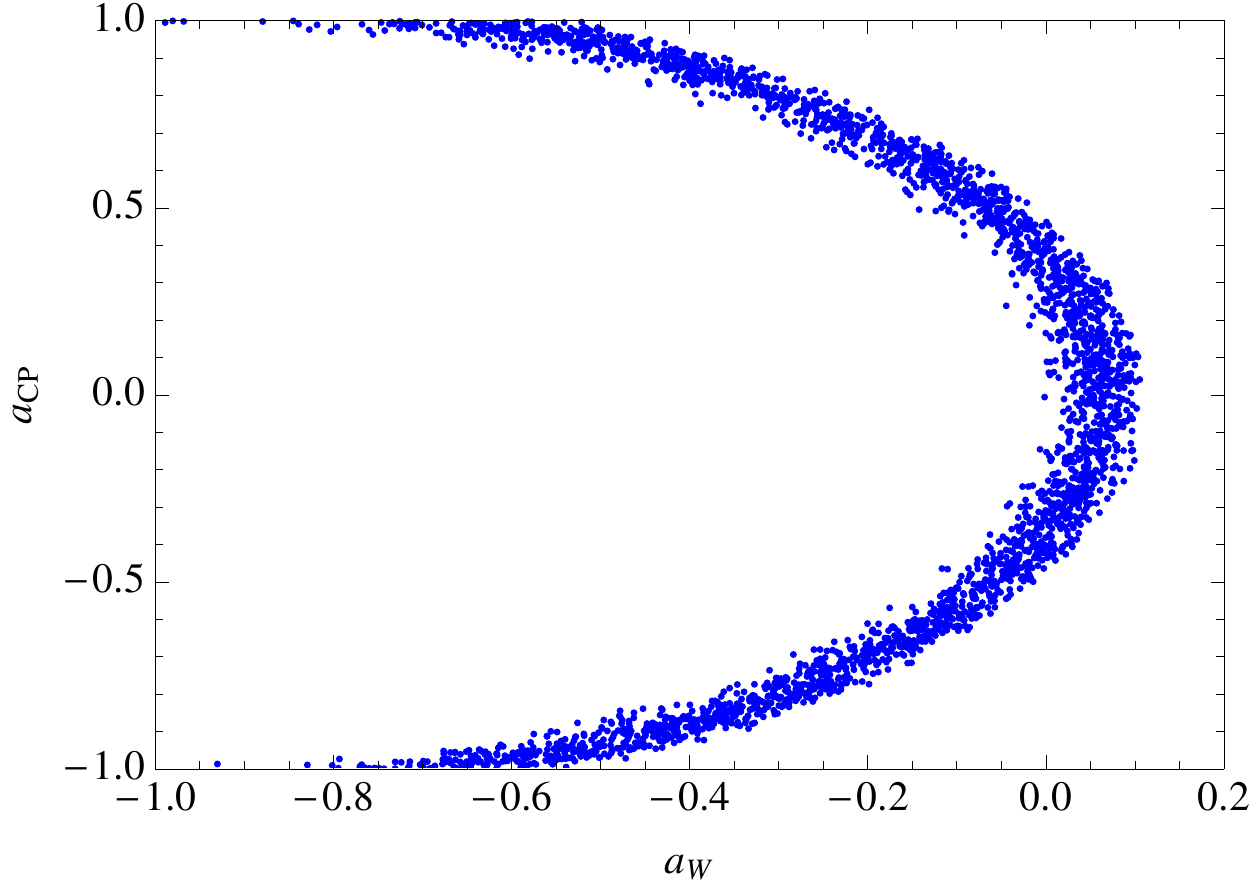}}
\caption{\it Results for the reference point $(|V_{ub}|,\,\gamma)=(3.5\times 10^{-3},\,66^\circ)$. Left panel: in red the SM prediction and its $1\sigma$ theoretical error bands for $\vep_K$ and $R_{BR/\Delta M}$ for this reference point; in orange (green) the $1\sigma$, $2\sigma$ and $3\sigma$ (from the darker to the lighter) experimental error ranges for $\vep_K$ ($R_{BR/\Delta M}$), in blue the correlation between $\vep_K$ and $R_{BR/\Delta M}$ induced by NP contributions. Right panel: allowed values for $a_W$ and $a_{CP}$ upon the setup of the left panel. See Ref.~\cite{Alonso:2012jc} for further details.}
\label{fig:EpsilonKRatio}
\end{figure}

Once  this reduced parameter space is identified, it is illustrative to choose one of its points as reference point, in order to present the features of this MFV scenario; for instance for the values $(|V_{ub}|,\gamma)=(3.5\times 10^{-3}, 66^\circ)$, $S_{\psi K_S}$, $R_{\Delta M_B}$ and $|V_{ub}|$ are all inside their own $1\sigma$ values, and the predicted SM values for $\epsilon_K$ and $R_{BR/\Delta M}$ are\footnote{The predicted SM value for $\epsilon_K$ differs from that in Ref.~\cite{Alonso:2012jc} due to the new input parameters used: in particular $\hat B_K=0.7643\pm0.0097$ has sensibly increased~\cite{Laiho:2009eu}.}
\beq
\epsilon_K=1.88\times10^{-3}\,,\qquad\qquad
R_{BR/\Delta M}=1.62\times 10^{-4}\,,
\eeq
that should be compared to the corresponding experimental determinations\footnote{$\left(R_{BR/\Delta M}\right)_{exp}$ has been computed considering the recent world average $BR(B^+\to\tau^+\nu)=\left(0.99\pm0.25\right)\times 10^{-4}$ from Ref.~\cite{UTFIT}.,
\beq
\left(\epsilon_K\right)_{exp}=\left(2.228\pm0.011\right)\times10^{-3}\,,\qquad\qquad
\left(R_{BR/\Delta M}\right)_{exp}=\left(1.95\pm0.49\right)\times 10^{-4}\,,
\eeq}
The errors on these quantities are  $\sim15\%$ and $\sim8\%$, estimated considering the uncertainties on the input parameters and the analysis performed in Ref.~\cite{Brod:2011ty}.
Fig.~\ref{fig:EpsilonKRatio} shows the correlation between $\epsilon_K$ and $R_{BR/\Delta M}$ (left panel) and the $a_{CP}-a_W$ parameter space (right panel), requiring that $\epsilon_K$ and $R_{BR/\Delta M}$ lie inside their own $3\sigma$ experimental determination. Finally, for those points in the $a_{CP}-a_W$ parameter space that pass all the previous constraints, the predictions for $S_{\psi\phi}$ and the $B$ semileptonic CP-asymmetry turned out to be close to the SM determination, in agreement with the recent LHCb measurements\cite{LHCb:2011aa}.

 In the next subsection, new constraints on  the $d=4$ operator coefficients $a_W$ and $a_{CP}$ will be obtained from their impact at loop-level on radiative $B$ decays. The latter data will be also used to constrain the set of $d=5$ chiral operators coefficients identified in Sect.~\ref{subsecd5}: while they are expected a priori to be all of comparable strength, the most powerful experimental constraints should result from the tree-level impact of  dipole operators $\mathcal X_1$ to $\mathcal X_8$, as they include vertices involving just three fields,  one of them being a light gauge boson.  Photonic penguins and also gluonic penguins and tree-level four-fermion diagrams (through renormalization group mixing effects) will be explored below and contrasted with radiative $B$ decays.

\boldmath
\subsection{$\bar{B}\to X_s\gamma$ branching ratio}
\unboldmath

The current experimental value of the $\bar{B}\to X_s\gamma$ branching ratio \cite{Lees:2012ufa} is
\begin{equation}
Br(\bar{B}\to X_s \gamma)=(3.31\pm 0.16 \pm 0.30\pm0.10)\times 10^{-4}\,,
\label{BRexp}
\end{equation}
for a photon-energy cut-off $E_\gamma>1.6$ GeV. On the other hand, its NNLO SM prediction  for that same energy cut-off  and in the $\bar{B}$-meson rest frame, 
reads\cite{Gambino:2001ew,Misiak:2006zs,Misiak:2006ab}
\begin{equation}
Br(\bar B\to X_s \gamma)=(3.15\pm0.23)\times 10^{-4}\,.
\label{BRSMNNLO}
\end{equation}
The presence of NP can easily modify this prediction, and the precision of both the experimental measure and the SM computation 
allows in principle to provide severe bounds on the NP parameters. 

The effective Lagrangian relevant for $b\to s\gamma$ decay at the $\mu_b=\mathcal{O}(m_b)$ 
scale can be written as:
\begin{equation}
\mathcal{L}_{eff} = \dfrac{4G_F}{\sqrt{2}}V^*_{ts}V_{tb}\left[\sum_{i=1}^{6}C_i(\mu_b)Q_i(\mu_b) \,+\, 
  C_{7\gamma}(\mu_b)Q_{7\gamma}(\mu_b) \,+\, C_{8G}(\mu_b)Q_{8G}(\mu_b)\right]\,,
\label{bseff_mb}
\end{equation}
where  $Q_{1,2}$, $Q_{3,\ldots,6}$ and $Q_{7\gamma,8G}$  denote the current-current, QCD penguin and magnetic dipole operators, respectively, as it is customary. In this effective Lagrangian, subleading terms proportional to $V^*_{us}V_{ub}$ have been neglected; the same applies to the contributions from the so-called primed operators, similar to those appearing in Eq.~(\ref{bseff_mb}) although with opposite chirality structure, which are suppressed by the $m_s/m_b$ ratio.
 
The value of the Wilson coefficients $C_i(\mu_b)$ at the scale $\mu_b$ is derived applying the QCD 
renormalisation group (RG) analysis to the corresponding Wilson coefficients, evaluated at the 
effective scale $\mu$ of the underlying theory, which is the matching scale linking the effective and full descriptions. For the SM case, this is the electroweak scale $\mu_W=\mathcal{O}(M_W)$. The effects of the RG contributions are in general non-negligible, and indeed the rate of the $b\to s\gamma$ decay in the SM is enhanced by a factor of $2-3$~\cite{Misiak:2006zs} upon the inclusion of these corrections. They originate dominantly from the mixing of charged current-current operators with the dipole operators, and to a smaller extent from the mixing with QCD-penguin operators. These QCD contributions can be formally written as 
\begin{equation}
C_i(\mu_b)=U_{ij}(\mu_b,\, \mu)\,C_j(\mu)\,,
\end{equation}
where $U_{ij}(\mu_b,\, \mu)$ are the elements of the RG evolution matrix from the effective scale $\mu$ down to $\mu_b$~\cite{Buchalla:1995vs}.  

The expression for the $\bar{B}\to X_s \gamma$ branching ratio is then given as follows:
\begin{equation}
Br(\bar{B}\to X_s \gamma) = R \left(|C_{7\gamma}(\mu_b)|^2+N(E_\gamma)\right),
\label{BRtotal}
\end{equation}
where $R=2.47\times10^{-3}$ is simply an overall factor as discussed in Refs.~\cite{Gambino:2001ew,Misiak:2006ab} 
and $N(E_\gamma)=(3.6\pm0.6)\times10^{-3}$ is a non-perturbative contribution for the photon-energy cut-off 
$E_\gamma>1.6$ GeV.  $C_{7\gamma}(\mu_b)$ can be decomposed into SM and NP contributions,
\begin{equation}
C_{7\gamma}(\mu_b)=C^{SM}_{7\gamma}(\mu_b)+\Delta C_{7\gamma}(\mu_b)\,, 
\label{eq:C7efftotal}
\end{equation}
where, for $\mu_b=2.5$ GeV, the SM contribution at the NNLO level, is given by~\cite{Gambino:2001ew,Misiak:2006zs,Misiak:2006ab}
\begin{equation}
C^{SM}_{7\gamma}(\mu_b)=-0.3523\,.
\label{eq:C7effSM}
\end{equation}
In our context, the NP contributions arise from the non-unitarity of the CKM matrix and the presence of flavour violating $Z$-fermion couplings (induced by the $d=4$ chiral operators $\cO_{1-4}$ \cite{Alonso:2012jc}), and from the direct contributions from the $d=5$ chiral operators $\cX_{1-8}$. In the following we will discuss separately these contributions.

\boldmath
\subsubsection{$d=4$ contributions}
\unboldmath 

The effective scale of the $d=4$ chiral operators is $f\geq v$, but no contributions to the Wilson coefficients relevant for $b\to s\gamma$ arise at scales above the electroweak one. As a result, the analysis of these contributions is alike to that in the SM, except for the fact that the NP operators modify the initial conditions at $\mu_W$.  The Wilson coefficients at the scale $\mu_W$ can be written as
\begin{equation}
C_i(\mu_W)=C_i^{SM}(\mu_W)+\Delta C^{d=4}_i(\mu_W)\,,
\label{eq:C7efftotal_muW}
\end{equation}
where the SM coefficients at the LO are given by\cite{Inami:1980fz}
\begin{equation}
\begin{aligned}
C_2^{SM}(\mu_W) &=1 \,,\\
C_{7\gamma}^{SM}(\mu_W) &=\frac{7x_t-5x_t^2-8x_t^3}{24(x_t-1)^3}+\frac{-2x_t^2+3x_t^3}{4(x_t-1)^4} \log x_t\,,\\
C_{8G}^{SM}(\mu_W) &=\frac{2 x_t+5 x_t^2- x_t^3}{8(x_t-1)^3}+\frac{-3 x_t^2}{4 (x_t-1)^4} \log x_t\,,
\end{aligned}
\label{SMcond}
\end{equation}
with $x_t\equiv m_t^2/M_W^2$. 

The NP contributions due to the non-unitarity of the CKM matrix induce modifications in all three Wilson 
coefficients involved:
\begin{equation}
\begin{aligned}
\Delta C_2^{d=4}(\mu_W)&=\left(a_W-i\,a_{CP}\right)y_b^2+\left(a_W^2+a_{CP}^2\right)y_b^2\,y_c^2\,,\\
\Delta C_{7\gamma}^{d=4}(\mu_W)&=\left(2 a_W y_t^2+\left(a_W^2+a_{CP}^2\right)y_t^4\right)
    \left(\frac{23}{36}+C_{7\gamma}^{SM}(\mu_W)\right)\,,\\
\Delta C_{8G}^{d=4}(\mu_W)&=\left(2 a_W y_t^2+\left(a_W^2+a_{CP}^2\right)y_t^4\right)
    \left(\frac{1}{3}+C_{8G}^{SM}(\mu_W)\right)\,.
\end{aligned}
\label{d4cond}
\end{equation}
These terms originate from the corresponding SM diagrams with the exchange of a $W$ boson and are 
proportional to $a_W$ and $a_{CP}$: indeed they are due to the modified vertex couplings, both in the 
tree-level diagram that originates $Q_2$ and in the 1-loop penguin diagrams that give rise to $Q_{7\gamma}$ 
and $Q_{8G}$. On the other hand, the new flavour-changing $Z$-fermion vertices participate in  penguin diagrams 
contributing to the $b\to s\gamma$ decay amplitude, with a $Z$ boson running in the loop~\cite{Buras:2011zb}. These 
contributions can be safely neglected, though, because they are proportional to the  $a_{Z}^{u,d}$ parameters, 
which are already severely constrained from their tree-level impact on other FCNC processes.

Including the QCD RG corrections, the NP contributions at LO to the Wilson coefficients are given by:
\begin{equation}  
\Delta C_{7\gamma}(\mu_b)=\eta^\frac{16}{23}\Delta C_{7\gamma}(\mu_W)+\dfrac{8}{3}\left(\eta^\frac{14}{23}
 -\eta^\frac{16}{23}\right)\Delta C_{8G}(\mu_W)+\Delta C_2(\mu_W)\sum_{i=1}^8\kappa_i\eta^{\sigma_i}\,,
\label{eq:DeltaC7eff}
\end{equation}
with
\begin{equation}  
\eta\equiv\dfrac{\alpha_s(\mu_W)}{\alpha_s(\mu_b)}=0.45\,.
\end{equation}
Here $\kappa$'s and $\sigma$'s are the magic numbers listed in Tab.~\ref{tab:c7magicnumbers}, while $\eta$ 
has been calculated taking $\alpha_s(M_Z=91.1876\,\text{GeV})=0.118$. Due to the simple additive structure 
of the NP contributions in Eq.~(\ref{eq:C7efftotal_muW}), these magic numbers are the same as in the SM context.

\begin{figure}[h!]
 \centering
\includegraphics[width=12cm]{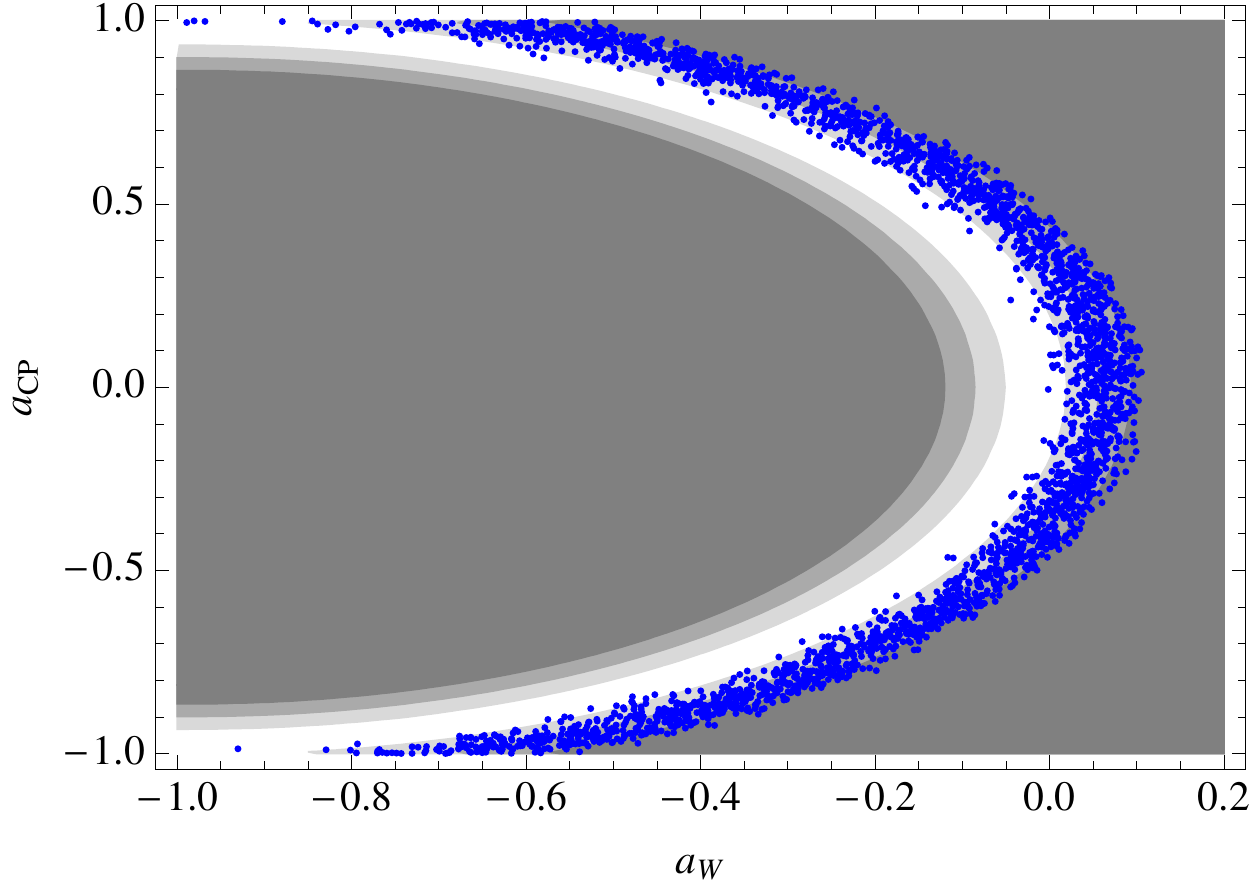}
\caption{\it $a_W-a_{CP}$ parameter space for $\varepsilon_K$ and $BR(B^+\to\sigma^+\nu)/\Delta M_{B_d}$ 
observables inside their $3\sigma$ error ranges and $a_Z^d\in{[}-0.044,0.009{]}$ (see \cite{Alonso:2012jc} for details). The gray areas correspond to the bounds from the $BR(\bar B\to X_s\gamma)$ at $1\sigma$, $2\sigma$, and $3\sigma$, from the lighter to the darker, respectively.}
\label{fig:EpsilonKRatioBSG}
\end{figure}

\begin{table}[h!]
\begin{center}
\begin{tabular}{|c||r|r|r|r|r|r|r|r|}
  \hline
  &&&&&&&&\\[-3mm]
  $i$ & $1$ & $2$ & $3$ & $4$ & $5$ & $6$ & $7$ & $8$ \\[1mm]
  \hline
  &&&&&&&&\\[-2mm]
  $\sigma_i$ & $\frac{14}{23}$ & $\frac{16}{23}$ & $\frac{6}{23}$ & $-\frac{12}{23}$ & $0.4086$ & $-0.4230$ 
           & $-0.8994$ & $0.1456$ \\[2mm]
  $\kappa_i$ & $2.2996$ & $-1.0880$ & $-\frac{3}{7}$ & $-\frac{1}{14}$ & $-0.6494$ & $-0.0380$ & $-0.0185$ 
             & $-0.0057$ \\[2mm]
  \hline
\end{tabular}
\caption{The magic numbers for $\Delta C_{7\gamma}(\mu_b)$ defined in Eq.~(\ref{eq:DeltaC7eff}).}
\label{tab:c7magicnumbers}
\end{center}
\end{table}
The analysis above allows to estimate the impact of the experimental value for $BR(\bar B\to X_s \gamma)$ on the NP parameter space of $\cO_1\dots \cO_4$ operators: in Fig.~\ref{fig:EpsilonKRatioBSG} we retake the scatter plot shown in Fig.~\ref{fig:EpsilonKRatio}b, based on the analysis of $\Delta F=1$ and $\Delta F=2$ observables for the reference point $(|V_{ub}|,\,\gamma)=(3.5\times 10^{-3},\,66^\circ)$, and superimpose  the new constraints resulting from the  present loop-level impact on $BR(\bar B\to X_s \gamma)$: they are depicted as shadowed (grey) exclusion regions. The figure illustrates that they reduce the available parameter space, eliminating about half of the points previously allowed in the scatter plot of Fig.~\ref{fig:EpsilonKRatio}b.

Fig.~\ref{fig:EpsilonKRatioBSG}  shows that $a_{CP}$, the overall coefficient of the genuinely CP-odd coupling $\cO_4$, and thus of $\cO_4(h)$ in Eq.~(\ref{defOh}), is still loosely constrained by low-energy data. This has an interesting phenomenological consequence on Higgs physics prospects, since it translates into correlated exotic Higgs-fermion couplings, which for instance at leading order in $h$ read: 
\beq
\delta \LL_{\chi=4}^h \supset a_{CP}\left(1 + \beta_{CP}\frac{h}{v}\right)\,\cO_4\,\,.
\eeq
For intermediate values of $\xi$ (for which the linear expansion could be an acceptable guideline), the relative weight of the couplings with and without an external Higgs particle reduces to -see Eq.~(\ref{d4OperatorsH-linear})-
\beq
\beta_{CP}\sim\, 4\,.
\eeq
These are encouraging results in the sense of allowing short-term observability. In  a conservative perspective, the operator coefficients of the $d=4$ non-linear expansion should be expected to be $\cO(1)$. Would this be the case, the possibility of NP detection would be delayed until both low-energy flavour experiments and LHC precision on  $h$-fermion couplings nears the $\cO(10^{-2})$ level, which for LHC means  to reach at least its $3000\,fb^{-1}$ running regime. Notwithstanding this, a steady improvement of the above bounds should be sought.

\boldmath
\subsubsection{$d=5$ contributions}
\unboldmath 

For the $d=5$ chiral operators considered, the effective scale weighting their overall strength is $\Lambda_s\leq 4 \pi f$. In the numerical analysis that follows, we will consider for $\Lambda_s$  the smallest value possible, i.e. $\Lambda_s=4\pi v$. For this value, the effects due to the $d=5$ chiral operators are maximized: indeed, for higher scales, the initial conditions for the Wilson coefficients are suppressed with the increasing of $\Lambda_s$. This effect is only slightly softened, but not cancelled, by the enhancement due to the QCD running from a higher initial scale. For the analytical expressions, we will keep the discussion at a more general level and  the high scale will be denoted by $\mu_s$, $\mu_s\gg v$. At this scale the top and $W$ bosons are still dynamical and therefore they do not contribute yet to any Wilson coefficients. The only operators relevant for $b\to s\gamma$ decay and  with non-vanishing initial conditions are thus $Q_{7\gamma}$ and $Q_{8G}$, whose contributions arise from dipole $d=5$ chiral 
operator. At the scale $\mu_s$  the Wilson coefficients can thus be written as
\beq
\begin{aligned}
C_i(\mu_s)&\equiv C_i^{SM}(\mu_s)+\Delta C_i^{d=5}(\mu_s)\,,\\
\end{aligned}
\eeq
where the only non-vanishing contributions are 
\begin{equation}
\Delta C_{7\gamma}^{d=5}(\mu_s)=d^d_F\frac{(4\pi)^2\, v\, y_t^2 }{\sqrt{2}\mu_s}\,,\qquad\qquad
\Delta C_{8G}^{d=5}(\mu_s)=d^d_G\frac{(4\pi)^2\, v\, y_t^2 }{\sqrt{2}\mu_s}\,, 
\end{equation}
 with $d^d_F$ and $d^d_G$ denoting the relevant photonic and gluonic dipole operator coefficients in Eq.~(\ref{LELag}), respectively.

The QCD RG analysis from $\mu_s$ down to $\mu_b$ should be performed in two distinct steps:
\begin{itemize}
\item[i)] A six flavour RG running from the scale $\mu_s$ down to $\mu_W$. Focusing on the Wilson coefficients corresponding to the SM and to the $d=5$ couplings under discussion, at the scale $\mu_W$ the coefficients read 
\beq
\begin{aligned}
C_i(\mu_W)&\equiv C_i^{SM}(\mu_W)+\Delta C_i^{d=5}(\mu_W)\,,
\label{WilsonMuWD5}
\end{aligned}
\eeq
where the only non-vanishing contributions from the set of $d=5$ flavour-changing fermionic operators are those given by
\beq
\begin{aligned}
C_{7\gamma}^{d=5}(\mu_W)&=\dfrac{8}{3}\left(1 - \eta_{\mu_s}^{2/21}\right) \eta_{\mu_s}^{2/3} \Delta C_{8G}^{d=5}(\mu_s)+ 
   \eta_{\mu_s}^{16/21}\Delta C_{7\gamma}^{d=5}(\mu_s) \,,\\
C_{8G}^{d=5}(\mu_W)&=\eta_{\mu_s}^{2/3}\Delta C_{8G}^{d=5}(\mu_s) \,, 
\end{aligned}
\label{d5cond}
\eeq
with
\begin{equation}  
\eta_{\mu_s}\equiv\dfrac{\alpha_s(\mu_s)}{\alpha_s(\mu_W)}\,.
\label{mm}
\end{equation}
In the numerical analysis $\eta_{\mu_s}=0.67$ will be taken.

\item[ii)] A five-flavour RG running from $\mu_W$ down to $\mu_b$. This analysis is alike to that described in the previous section, substituting  the initial conditions for the Wilson coefficients in Eq.~(\ref{eq:C7efftotal_muW})-Eq.~(\ref{d4cond}) for those in Eqs.~(\ref{WilsonMuWD5})-(\ref{mm}).
\end{itemize}

\begin{figure}[h!]
 \centering
 \subfigure[$BR(\bar B\to X_s\gamma)$ vs. $b^d_{eff}$]{
\includegraphics[width=7.8cm]{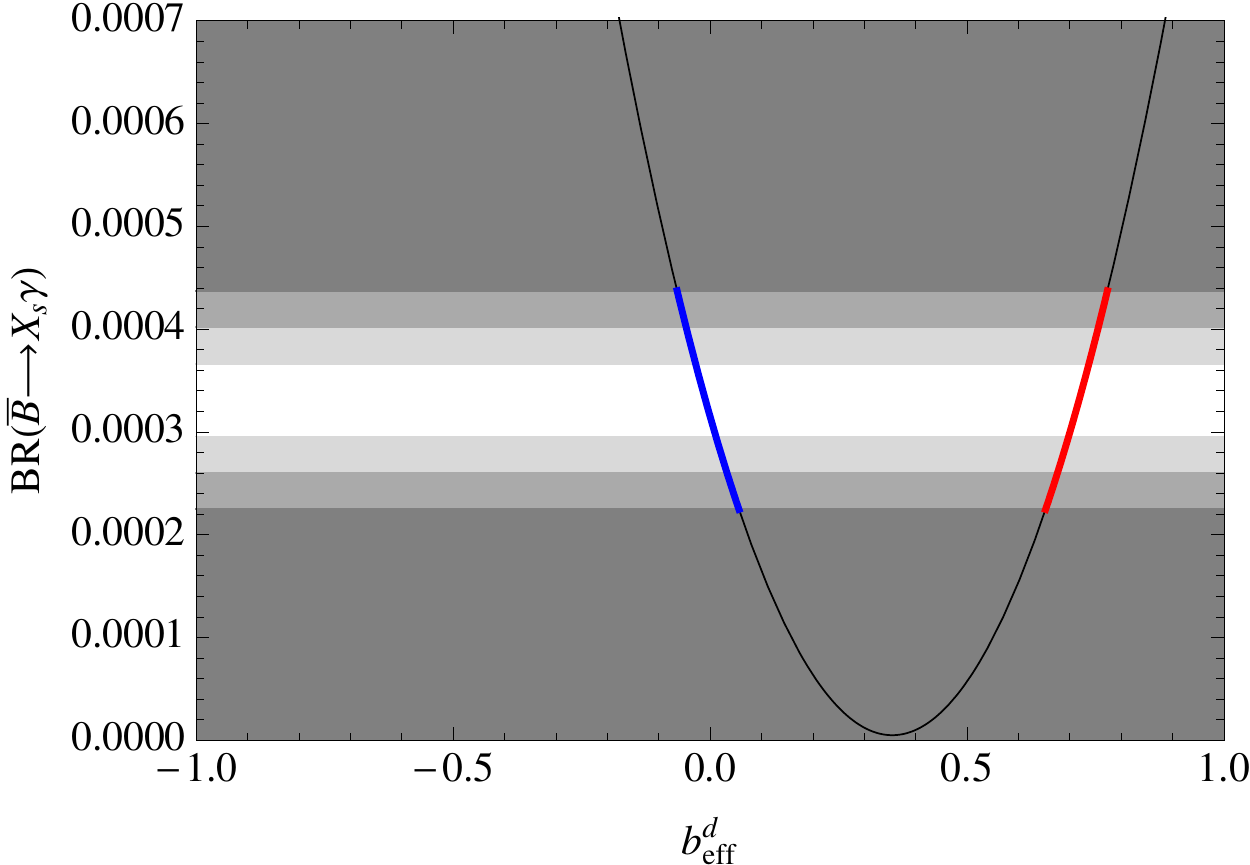}}  
\subfigure[$b^d_F-b^d_G$ parameter space]{
\includegraphics[width=7.6cm]{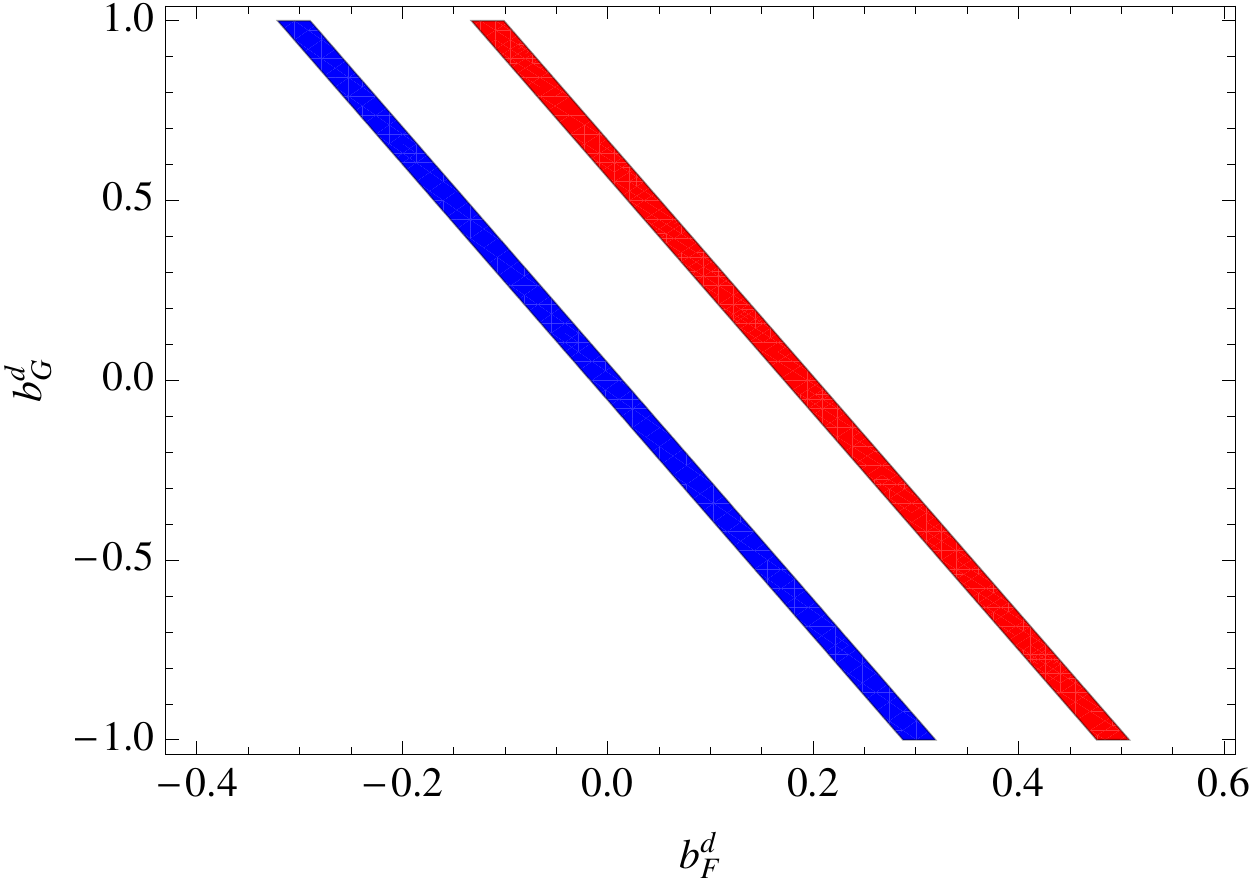}}
\caption{\it Left panel: the curve depicts  $BR(\bar B\to X_s\gamma)$ as a function of $b^d_{eff}$, while the horizontal bands are the experimentally excluded regions at $1\sigma$, $2\sigma$, and $3\sigma$, from the lighter to the darker, respectively. Right panel: the $3\sigma$ corresponding allowed $b^d_F-b^d_G$ parameter space is depicted as two separate narrow bands.}
\label{fig:BSG5f4pv}
\end{figure}

It is interesting to focus on the final numerical result for the $BR(\bar B\to X_s\gamma)$, leaving unspecified only the parameters of the $d=5$ chiral operators $b^d_{F,G}$:
\beq
BR(b\to s\gamma)=0.000315 - 0.00175\, b^d_{eff} + 0.00247\, \left(b^d_{eff}\right)^2\,,
\eeq
where
\beq
b^d_{eff}\equiv3.8\, b^d_F + 1.2\, b^d_G\,.
\label{befff4pv}
\eeq
The corresponding plot is shown on the left-hand side of Fig.~\ref{fig:BSG5f4pv}, which depicts the dependence of the branching ratio on $b^d_{eff}$, together with the experimental $3\sigma$ regions. Two distinct ranges for $b^d_{eff}$ are allowed:
\beq
-0.07 \lesssim b^d_{eff} \lesssim 0.04\qquad
\text{or}\qquad 
0.67 \lesssim b^d_{eff} \lesssim 0.78\,.
\eeq
Using the expression for $b^d_{eff}$ in Eq.~(\ref{befff4pv}), it is possible to translate these bounds onto the $b^d_F-b^d_G$ parameter space, as shown on the right-hand side of Fig.~\ref{fig:BSG5f4pv}. The two narrow bands depict the two allowed regions. 

Analogously to the case of $\cO_1(h)\dots \cO_4(h)$ operators discussed in the previous subsection, a correlation would hold between a low-energy signal from these $d=5$ couplings and the detection of exotic fermionic couplings at LHC, upon considering their extension  to include $h$-dependent insertions. Nevertheless, a consistent analysis would require in this case to consider $d=6$ couplings of the non-linear expansion, which are outside the scope of the present work.

%
%

\section{Conclusions}
\label{Conclusions}

The lack of indications of new resonances at LHC data other than a strong candidate to be the SM scalar boson $h$, together with the alignment of the couplings of the latter with SM expectations, draws a puzzling panorama with respect to the electroweak hierarchy problem.  If the experimental  pattern persists, either the extended prejudice against fine-tunings of the SM parameters should be dropped, or the new physics scale is still awaiting discovery and may be associated  for instance to a dynamical origin of the SM scalar boson. We have focused in this work on possible implications for fermionic couplings of a strong interacting origin of electroweak symmetry breaking dynamics with a light  scalar $h$ with mass around $125$ GeV, within an effective Lagrangian approach. 

The parameter describing the degree of non-linearity $\xi=(v/f)^2$ must lie in the range $0<\xi<1$. For small values, the effective theory converges towards the SM, as the NP contributions can be safely neglected. On the other hand, large values indicate a chiral regime for the dynamics of the Goldstone bosons, which in turn requires to use a chiral expansion to describe them, combined with appropriate insertions of the light $h$ field. 

We identified the flavour-changing operator basis for the non-linear regime up to $d=5$. Furthermore, taking into account the QCD RG evolution, the coefficients of these operators have been constrained from a plethora of low-energy transitions. In particular we have analyzed in detail and in depth the constraints resulting from the data on $\bar{B}\to X_s\gamma$ branching ratio. Its impact is important on the global coefficients of the four relevant $d=4$ flavour-changing chiral couplings at the loop level, and on those of the $d=5$ dipole operators. The limits obtained constrain in turn the possible fermion-$h$ exotic couplings to be explored at the LHC. A particularly interesting example is that of the intrinsically CP-odd $d=4$ operator $\cO_4$ of the non-linear expansion, whose coefficient is loosely constrained by data: a correlation is established between the possible signals in low-energy searches of CP-violation and anomalous $h$-fermion couplings at the LHC. Their relative strength is explored for the case of a relatively small $\xi$. A similar correlation between low-energy flavour searches and LHC signals  also follows for all other operators.

\section*{Acknowledgements}
We acknowledge partial support by European Union FP7 ITN INVISIBLES (Marie Curie 
Actions, PITN- GA-2011- 289442), CiCYT through the project FPA2009-09017, CAM through the project 
HEPHACOS P-ESP-00346, European Union FP7 ITN UNILHC (Marie Curie Actions, PITN-GA- 2009-237920), 
MICINN through the grant BES-2010-037869 and the Juan de la Cierva programme (JCI-2011-09244), ERC Advance Grant ``FLAVOUR'' (267104), Te\-ch\-ni\-sche Universit\"at M\"unchen - Institute for 
Advanced Study - funded by the German Excellence Initiative, Italian Ministero dell'Uni\-ver\-si\-t\`a 
e della Ricerca Scientifica through the COFIN program (PRIN 2008) and the contracts MRTN-CT-2006-035505 
and  PITN-GA-2009-237920 (UNILHC). We thank the Galileo Galilei Institute for Theoretical Physics for 
the hospitality and the INFN for partial support during the completion of this work. R.A. acknowledges the Harvard Physics department for hospitality during the completion phase of this work. S.R. and J.Y acknowledge CERN TH department for hospitality during the completion phase of the work.

\appendix

%
%
\section{Relation to the SILH basis}
\label{AppA}

In this appendix we revisit the transition from an $SU(2)_L\times U(1)_Y$ invariant effective Lagrangian in the linear realization of the EW symmetry breaking mechanism to an effective chiral Lagrangian, focussing to the so-called SILH framework \cite{Giudice:2007fh}. The $d=6$ SILH Lagrangian in Eq.~(2.15) of Ref.~\cite{Giudice:2007fh} can be written in terms of $\UH$, $\VL$, $\TL$ and a scalar field $h$:
\bea
{\cal L}_{SILH} & = & \xi \left\{ 
   \frac{c_H}{2} (\partial_\mu h) (\partial^\mu h) \, \cF(h) \, + \, 
   \frac{c_T}{2} \frac{v^2}{4} \Tr\left[\TL\VL^\mu\right]\Tr\left[\TL\VL_\mu\right] \cF(h)^2 \right. + \nn \\
 & & \hspace{0.2 cm}  \left. - \,c_6 \lambda \frac{v^4}{8} \cF(h)^3 + 
  \left(c_y\frac{v}{2\sqrt{2}}\bar{Q}_L\UH\, \,{\rm diag}({\bf y_U},{\bf y_D})\, Q_R \, \cF(h)^{3/2} +\mbox{h.c.}\right)\right. + \nn\\
 & & \hspace{0.2 cm}  \left. - \,
  i\frac{c_W g}{2m^2_\rho} \frac{f^2}{2} \left(\cD_\mu W^{\mu \nu}\right)_i \Tr\left[\sigma_i\,\VL_{\!\nu}\right]
     \,\cF(h) \,+\, 
  i\frac{c_B g'}{2m^2_\rho} \frac{f^2}{2} \left(\partial_\mu B^{\mu \nu}\right) \Tr\left[\TL\,\VL_{\!\nu}\right]
     \,\cF(h) \right. + \nn \\
 & & \hspace{0.2 cm} \left. + \,
  i \frac{c_{HW} g}{16 \pi^2} W^{\mu \nu}_i \Big(\frac14 \Tr \left[\sigma_i\,\VL_{\!\mu}\VL_{\!\nu}\right]\,
   \cF(h) - \frac14 \Tr \left[\sigma_i\,\VL_{\!\mu}\right]\,\partial_\nu \cF(h) \Big) \right. + \nn \\   
 & & \hspace{0.2 cm} \left. + \,
  i \frac{c_{HB} g'}{16 \pi^2} B^{\mu \nu}\Big(\frac14 \Tr \left[ \TL\, \VL_{\!\mu}\VL_{\!\nu} \right] \, 
    \cF(h) + \frac14 \Tr \left[\TL\,\VL_{\!\mu} \right]\,\partial_\nu \cF(h) \Big) \right. \nn + \\ 
 & & \hspace{0.2 cm} + \, 
  \frac{c_\gamma {g'}^{2}}{16 \pi^2} \frac{g^2}{g^2_\rho} \frac{1}{2} B_{\mu \nu} B^{\mu \nu}\,\cF(h) + 
  \frac{c_g g_S^2}{16 \pi^2} \frac{y_t^2}{g^2_\rho} \frac{1}{2} G^a_{\mu \nu} G^{a \mu \nu}\,\cF(h) 
\bigg\} \, ,
\label{Frameworkrangian}
\eea
where the notation of the operator coefficients is as in Ref.~\cite{Giudice:2007fh} and $\cF(h)=F(h)$ is the function 
of the light Higgs fields resulting from the doublet Higgs ansatz as in Eq.~(\ref{Fs}); the Lagrangian above is only complete at leading order for values of $\xi\ll1$; otherwise other operators of non-linear parenthood have to be added, as earlier explained.

%
%
\section{Gauge fields equations of motion}
\label{AppB}

When deriving the gauge field SM equations in Eqs.~(\ref{triplet}) and (\ref{singlet}), all contributions from $d=4$ operators  in $\delta\LL_{d=4}$ effective Lagrangians have been neglected, on the assumption that their coefficients are small, typically $a_i<1$, $i=1...4$, with typically $a_i \approx 1/(16\pi^2)$. This allows to trade flavour-conserving currents for gauge terms with derivatives of the gauge field strengths.

Otherwise, for $a_i\sim 1$, taking into account that the gauge sector is already severely modified, and thus keeping only the flavour-changing contributions in $\delta\LL^f_{d=4}$, Eqs.~(\ref{triplet}) and (\ref{singlet}) would get modified to 
\begin{align}
\begin{split}
\left(D^\mu\,W_{\mu\nu}\right)_j =
& +\,i\,\dfrac{g}{4}\,v^2\,\Tr[\VL_\nu\,\sigma_j]\,\cF_C(h)+\dfrac{g}{2}\,\bar Q_L\,\gamma_\nu\,\sigma_j\,Q_L+\\ 
&-\frac{g}{2}\, \bar Q_L\,\gamma_\nu\, \lambda_F\,\left[ \left(a_2-a_3\right) \delta_{jk} - a_4\,  \epsilon_{3jk} \right] \, \sigma_k\,Q_L\,,
\label{tripletFV}
\end{split}\\
\begin{split}
\derp^\mu\,B_{\mu\nu}=
& -\,i\,\dfrac{g'}{4}\,v^2\,\Tr[\TL\,\VL_\nu]\,\cF_C(h) \,+\, g'\,\bar Q_L\,\gamma_\nu\,{\bf h}^L_q \,Q_L+g'\,\bar Q_L\,\gamma_\nu\,{\bf h}^R_q \,Q_L+\\
&-g'\,\bar Q_L\,\gamma_\nu\,\lambda_F \left[a_1 \unity + (a_2+a_3) \frac{\sigma_3 }{2}\right] \,Q_L\,,
\end{split}
\label{singletFV}
\end{align}
with ${\mathbf h}_{L,R}$ the left and right hypercharges in the $2\times 2$ matrix notation and where the right-handed flavour-changing contributions have been disregarded.  However, as gauge-$h$ coefficients are severely constrained by EW precision data (barring extremely fine-tuned regions in the parameter space) the analysis of flavour-changing couplings would get modified only by terms of $O(a_{i} \times a_{GH})$, $i=1...4$, being $a_{GH}$ the coefficients in the gauge-$h$ sector, and therefore their impact on the flavour sector is negligible.

%
%
\boldmath
\section{Linear siblings of the $d=5$ operators}
\label{AppC}
\unboldmath

In this appendix we connect the operators listed in Eqs.~(\ref{OpGroup3})-(\ref{OpGroup1}) with those defined 
in the linear realization, $\mathcal{X}_{i}\leftrightarrow\sum_j \mathscr{C}_{ij}\mathcal{X}_{Hj}$ where 
$\mathscr{C}$ is a $18\times18$ matrix.

The first set of non-linear operators listed in Eq.~(\ref{OpGroup3}) corresponds to the following eight linear 
operators containing fermions, the Higgs doublet $H$, the rank-2 antisymmetric tensor $\sigma^{\mu\nu}$ and the field 
strengths $B_{\mu\nu}$, $W_{\mu\nu}$ and $G_{\mu\nu}$: 
\beq
\begin{aligned}
&\mathcal{X}_{H1}=g'\,\bar Q_L\, \sigma^{\mu\nu}\,H\, D_R\,B_{\mu\nu}\\
&\mathcal{X}_{H2}=g'\,\bar Q_L\, \sigma^{\mu\nu}\,\tilde H\, U_R\,B_{\mu\nu}\\
&\mathcal{X}_{H3}=g\,\bar Q_L\, \sigma^{\mu\nu}\,W_{\mu\nu}\,H\, D_R\\
&\mathcal{X}_{H4}=g\,\bar Q_L\, \sigma^{\mu\nu}\,W_{\mu\nu}\,\tilde H\, U_R\\
&\mathcal{X}_{H5}=g_s\,\bar Q_L\, \sigma^{\mu\nu}\,H\, D_R\,G_{\mu\nu}\\
&\mathcal{X}_{H6}=g_s\,\bar Q_L\, \sigma^{\mu\nu}\,\tilde H\, U_R\,G_{\mu\nu}\,, \\
&\mathcal{X}_{H7}=g\,\bar Q_L\, \sigma^{\mu\nu}\,\sigma_i\,H\, D_R\,H^\dag\,W_{\mu\nu}\,\sigma^i\,H\\
&\mathcal{X}_{H8}=g\,\bar Q_L\, \sigma^{\mu\nu}\,\sigma_i\,\tilde H\, U_R\,H^\dag\,\sigma^i\,W_{\mu\nu}\,H \,.
\end{aligned}
\eeq
The operators $\mathcal{X}_{H7,H8}$ have mass dimension $d=8$, while all the others have (linear) mass 
dimension $d=6$. The correspondence among these linear operators and those non-linear listed in 
Eq.~(\ref{OpGroup3}) is the following: for $i=1,\ldots,8$,
\beq
\begin{gathered}
\mathcal{X}_{i}\leftrightarrow\sum_{j=1}^{8}\mathscr{C}_{ij}\mathcal{X}_{Hj}\qquad\qquad\text{with}\\
\mathscr{C}=\frac{\sqrt{2}}{f}
\left(
\begin{array}{cccccccc}
 1 & 1 & 0 & 0 & 0 & 0 & 0 & 0 \\
 -1 & 1 & 0 & 0 & 0 & 0 & 0 & 0 \\
 0 & 0 & 1 & 1 & 0 & 0& 0 & 0 \\
 0 & 0 & -1 & 1 & 0 & 0 & 0 & 0 \\
 0 & 0 & 0 & 0 & 1 & 1 & 0 & 0 \\
 0 & 0 & 0 & 0 & -1 & 1 & 0 & 0  \\
 0 & 0 & 1 & -1 & 0 & 0 &-4/f^2 & 4/f^2  \\
 0 & 0 &-1 & -1 & 0 & 0 & 4/f^2 & 4/f^2
\end{array}
\right)
\end{gathered}
\end{equation}

The second set of non-linear operators listed in Eq.~(\ref{OpGroup2}) corresponds to the following four linear 
operators containing fermions, the Higgs doublet $H$ and the rank-2 antisymmetric tensor $\sigma^{\mu\nu}$: 
\beq
\begin{aligned}
\mathcal{X}_{H9}&=\bar Q_L\, \sigma^{\mu\nu}\,H\, D_R\,\left(\left(D_\mu H\right)^\dagger  D_\nu H - (\mu\leftrightarrow \nu)\right)\,,\\
\mathcal{X}_{H10}&=\bar Q_L\, \sigma^{\mu\nu}\,\tilde H\, U_R\,\left(\left(D_\mu H\right)^\dagger  D_\nu H - (\mu\leftrightarrow \nu)\right)\,,\\
\mathcal{X}_{H11}&=\bar Q_L\,\sigma_i\, \sigma^{\mu\nu}\,H\, D_R\,\left(\left(D_\mu H\right)^\dagger \sigma^i D_\nu H - (\mu\leftrightarrow \nu)\right)\,,\\
\mathcal{X}_{H12}&=\bar Q_L\,\sigma_i\, \sigma^{\mu\nu}\,\tilde H\, U_R\,\left(\left(D_\mu H\right)^\dagger \sigma^i D_\nu H - (\mu\leftrightarrow \nu)\right)\,,\\
\end{aligned}
\eeq
all of them of mass dimension $d=8$. The correspondence among these linear operators and those non-linear listed in Eq.~(\ref{OpGroup2}) is the following: for $i=9,\ldots,12$,
\beq
\mathcal{X}_{i}\leftrightarrow\sum_{j=9}^{12}\mathscr{C}_{ij}\mathcal{X}_{Hj}\qquad\text{with}\qquad
\mathscr{C}=\frac{2\sqrt{2}}{f^3}
\left(
\begin{array}{cccc}
 0 & 0 & 1 & 1 \\
 0 & 0 & -1 & 1 \\
 1 & -1 & 0 & 0 \\
 -1 & -1 & 0 & 0
\end{array}
\right)
\end{equation}

For the third set in Eq.~(\ref{OpGroup1}), we consider the following six linear operators involving fermions 
and the Higgs doublet $H$:
\beq
\begin{aligned}
\mathcal{X}_{H13}&=\bar{Q}_L\, H\, D_R\, \left(D_\mu H \right)^\dagger D^\mu H\,, \\
\mathcal{X}_{H14}&=\bar{Q}_L\, \tilde H\, U_R\, \left(D_\mu H \right)^\dagger D^\mu H\,,\\
\mathcal{X}_{H15}&=\bar{Q}_L\, \sigma_i\,H\, D_R\, \left(D_\mu H \right)^\dagger\sigma^i D^\mu H\,,\\
\mathcal{X}_{H16}&=\bar{Q}_L\, \sigma_i\,\tilde H\, U_R\, \left(D_\mu H \right)^\dagger\sigma^i D^\mu H\,,\\
\mathcal{X}_{H17}&=\bar{Q}_L\, H\, D_R\, \left(D_\mu H \right)^\dagger H\, H^\dagger D^\mu H\,,\\
\mathcal{X}_{H18}&=\bar{Q}_L\, \tilde H\, U_R\, \left(D_\mu H \right)^\dagger H\, H^\dagger D^\mu H\,,
\end{aligned}
\eeq
where the first four operators have mass dimension $d=8$, while the last two have mass dimension $d=10$. It is then 
possible to establish the following correspondence between these linear operators and those non-linear listed 
in Eq.~(\ref{OpGroup1}): for $i=13,\ldots,18$,
\beq
\mathcal{X}_{i}\leftrightarrow\sum_{j=13}^{18} \mathscr{C}_{ij}\mathcal{X}_{Hj}\qquad\text{with}\qquad
\mathscr{C}=\frac{2\sqrt{2}}{f^3}\left(
\begin{array}{cccccc}
 -1 & -1 & 0 & 0 & 0 & 0 \\
 1 	& -1 & 0 & 0 & 0 & 0 \\
 0 & 0 & 1 & 1 & 0 & 0 \\
 0 & 0 & -1 & 1 & 0 & 0 \\
 2 & 2 & 1 & -1 & -8/f^2 & -8/f^2 \\
 -2 & 2 & -1 & -1 & 8/f^2 & -8/f^2
\end{array}
\right)\,.
\eeq

%
%
\boldmath
\section{$d=5$ operator coefficients in the unitary basis}
\label{AppD}
\unboldmath

In this appendix, we report the relations between the coefficients appearing in the Lagrangian Eq.~(\ref{LELag}) and the ones defined in Eq.~(\ref{OrigLag5}) for the effective Lagrangian in the unitary basis: 
\beq
\left(
\begin{array}{c}
c_W^u\\
c^d_W\\
c^+_{WZ}\\
c^-_{WZ}\\
d^u_F\\
d^d_F\\
d^u_Z\\
d^d_Z\\
d^+_W\\
d^-_W\\
d_G^u\\
d_G^d
\end{array}\right)=\mathcal{A}
\left(
\begin{array}{c}
b_1\\
\\
\\
\\
\\
\\
\cdots\\
\\
\\
\\
\\
b_{12}
\end{array}\right)\,,\qquad\qquad
\left(\begin{array}{c}
b^u_Z\\
b^d_Z\\
b^u_W\\
b^d_W\\
b^+_{WZ}\\
b^-_{WZ}\\
\end{array}\right)
=\mathcal{B}
\left(
\begin{array}{c}
b_{13}\\
\\
\\
\cdots\\
\\
b_{18}
\end{array}\right)
\eeq
\beq
\mathcal{A}=\left(
\begin{array}{cccccccccccc}
 0 & 0 & 2 i & 2 i & 0 & 0 & 2 i & 2 i & -1 & -1 & 1 & 1 \\
 0 & 0 & -2 i & 2 i & 0 & 0 & 2 i & -2 i & 1 & -1 & -1 & 1 \\
 0 & 0 & 0 & 0 & 0 & 0 & 0 & 0 & 4 & -4 & 0 & 0 \\
 0 & 0 & 0 & 0 & 0 & 0 & 0 & 0 & -4 & -4 & 0 & 0 \\
 1 & 1 & 1 & 1 & 0 & 0 & 1 & 1 & 0 & 0 & 0 & 0 \\
 1 & -1 & -1 & 1 & 0 & 0 & 1 & -1 & 0 & 0 & 0 & 0 \\
 -2 s_W^2 & -2 s_W^2 & 2 c_W^2 & 2 c_W^2 & 0 & 0 & 2 c_W^2 & 2 c_W^2 & 0 & 0 & 0 & 0 \\
 -2 s_W^2 & 2 s_W^2 & -2 c_W^2 & 2 c_W^2 & 0 & 0 & 2 c_W^2 & -2 c_W^2 & 0 & 0 & 0 & 0 \\
 0 & 0 & 2 & -2 & 0 & 0 & 2 & -2 & 0 & 0 & 0 & 0 \\
 0 & 0 & 2 & 2 & 0 & 0 & -2 & -2 & 0 & 0 & 0 & 0 \\
 0 & 0 & 0 & 0 & 1 & 1 & 0 & 0 & 0 & 0 & 0 & 0 \\
 0 & 0 & 0 & 0 & 1 & -1 & 0 & 0 & 0 & 0 & 0 & 0
\end{array}
\right)
\label{bCoeff}
\eeq

\beq
\mathcal{B}=\left(
\begin{array}{cccccc}
 -1 & -1 & -1 & -1 & -1 & -1 \\
 -1 & 1 & 1 & -1 & -1 & 1 \\
 -1 & -1 & 1 & 1 & 1 & 1 \\
 -1 & 1 & -1 & 1 & 1 & -1 \\
 0 & 0 & -2 & 2 & -2 & 2 \\
 0 & 0 & -2 & -2 & 2 & 2
\end{array}
\right)
\eeq



\providecommand{\href}[2]{#2}\begingroup\raggedright\endgroup

\end{document}